\documentclass[pdftex,twocolumn,epjc3]{svjour3}
\smartqed  
\RequirePackage{graphicx}
\RequirePackage[T1]{fontenc}
\usepackage{sidecap}
\usepackage{subfigure}
\usepackage{float}
\usepackage{xcolor}

\usepackage{hepnames}
\usepackage{hepnicenames}
\usepackage{xpatch}
\makeatletter
\xpatchcmd\@HepConStyle
{\edef\@upcode{\updefault}}
{\ifdefined\shapedefault\edef\@upcode{\shapedefault}\else\edef\@upcode{\updefault}\fi}
{}{}
\makeatother

\newcommand{\dEdx}{\ensuremath{\mathrm{d}E/\mathrm{d}x}}

\usepackage{soul}

\RequirePackage{latexsym}
\RequirePackage[numbers,sort&compress]{natbib}
\RequirePackage[colorlinks,citecolor=blue,urlcolor=blue,linkcolor=blue]{hyperref}
\journalname{Eur. Phys. J. C}
\begin{document}

\title{Cluster time measurement with CEPC calorimeter
}


\author{Yuzhi Che\thanksref{addr1,addr2}
  \and
  Vincent Boudry\thanksref{addr3}
  \and
  Henri Videau \thanksref{addr3}
  \and
  Muchen He\thanksref{addr1}
  \and
  Manqi Ruan\thanksref{e1,addr1}
}

\thankstext{e1}{e-mail: ruanmq@ihep.ac.cn}

\institute{Institute of High Energy Physics, Chinese Academy of Sciences, Beijing 100049, China \label{addr1}
  \and
  University of Chinese Academy of Sciences, 19A Yuquan Road, Beijing 100049, China \label{addr2}
  \and
  LLR, Ecole Polytechnique, Palaiseau, France \label{addr3}
}

\date{Received: date / Accepted: date}

\maketitle

\begin{abstract}
  We have developed an algorithm dedicated to timing reconstruction in highly granular
  calorimeters(HGC). The performance of this algorithm is evaluated on an electromagnetic
  calorimeter (ECAL) with geometries comparable to the electromagnetic compartment (CE-E)
  of the CMS endcap calorimeter upgrade at HL-LHC and conceptual Particle Flow oriented ECAL's
  for future Higgs factories. The time response of individual channel is parameterized according
  to the CMS experimental result~\cite{akchurin2017timing}. The particle Time-of-Flight (ToF) can
  be measured with a resolution of $5\sim20 \;\rm{ps}$ for electromagnetic (EM) showers and
  $80\sim 160 \;\rm{ps}$ for hadronic showers above 1 GeV. The presented algorithm provides
  comparable reconstruction with the $E_{\mathrm{hit}}^2$ weighting strategy and can
  significantly improve the time resolution compared to a simple averaging of the fast component
  of the time spectrum. The effects of three detector configurations are also quantified in this
  study. ToF resolution depends linearly on the timing resolution of a single silicon sensor and
  improves statistically with increasing incident particle energy. The timing layers at depth of
  $6\sim 9$ radiation lengths provide higher timing performance for EM showers. A clustering
  algorithm that vetoes isolated hits improves ToF resolution.
\end{abstract}

\section{Introduction}\label{sec1}

Precise Time-of-Flight (ToF) reconstruction is important for experiments in
particle physics at the high energy frontier. As the world's most powerful
particle accelerator, the Large Hadron Collider (LHC) is expected to deliver
proton-proton collisions with an integrated luminosity of $300 \rm{~fb^{-1}}$
by the end of 2023. From 2026 to about 2030, this machine will be upgraded into
the High Luminosity configuration (HL-LHC) \cite{apollinari2017high} and
collect $3000 \rm{~fb^{-1}}$ more data. With a tenfold increase in luminosity,
the corresponding number of collisions per bunch crossing (a.k.a. pile-up) is
expected to be $140\sim 200$, $5\sim 7$ times the value of the LHC. The event
selection and characterisation at the HL-LHC will face the increasing
difficulty of assigning the detector signals to the correct interaction. Since
the typical time spread of pile-up events is at the hundred picosecond level, a
ToF measurement with a resolution of about $20 \sim 30 \rm{~ps}$ can
significantly mitigate the effect of pile-up
\cite{bornheim2015precision,cms2017technical,collaboration2017phase,cerri2018cms,atlas2020technical}.

For future electron-positron colliders, the $\Pelectron\Ppositron$ Higgs
factories are identified as the highest-priority next collider by the European
Strategy statement \cite{european20202020}. As one of the collider concepts,
the circular $\Pelectron\Ppositron$ collider can also operate at a
center-of-mass energy of 91.2 GeV for a Z factory with high luminosity,
providing a valuable flavor physics opportunity. Particle identification (PID)
is critical for flavor physics measurements. A common method for separating
$\PK/\Ppi/\Pp$ is to measure the ToF and \dEdx ~of the particles. The
$\PK/\Ppi$ and $\PK/\Pp$ separation power provided by \dEdx ~decreases sharply
as the particle momentum approaches 1 GeV and 2 GeV. Therefore, the ToF plays
an essential role in compensating for the lack of PID performance provided by
\dEdx. For instance, a ToF precision better than 50 ps makes significant PID
improvement at the CEPC \cite{an2018monte}.

\begin{figure}[htbp]
  \centering
  \includegraphics[width=0.43\textwidth]{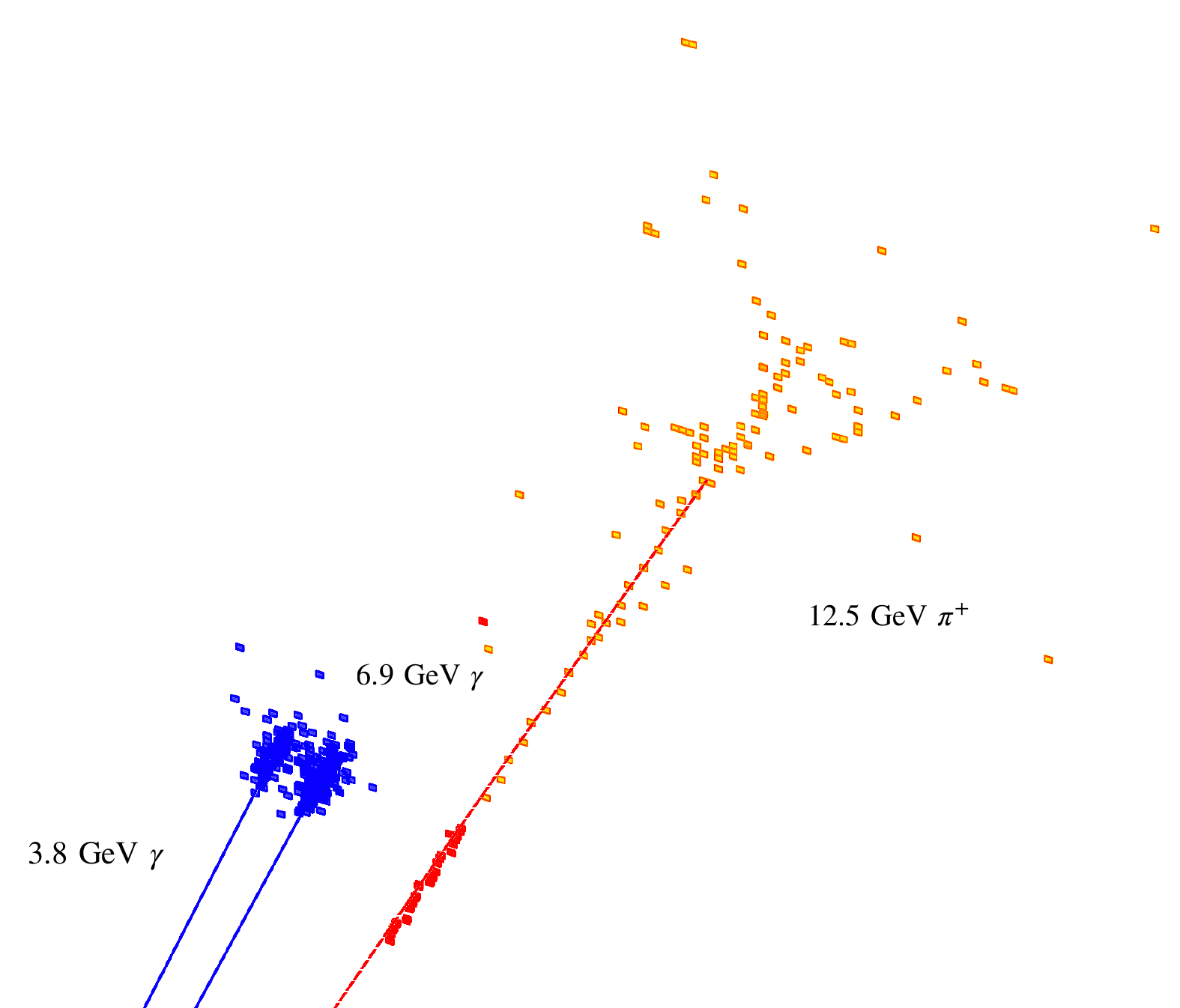}

  \caption{The event display of two photons and one charged pion, generated from a
    hadronic $\Ptau$ decay. The blue hits around the blue solid lines are the ECAL hits of
    a 3.8 GeV photon shower and a 6.9 GeV photon shower. The red hits around the red
    dashed line are the ECAL hits belonging to a 12.5 GeV $\Ppiplus$ shower, while the
    orange hits are the HCAL hits of the same $\Ppiplus$ shower. The size of these cells
    is $1\times 1 \rm{~cm}$.}
  \label{display}
\end{figure}

The concept of high granularity calorimetry concept is widely applied in the
upgrade detectors for the HL-LHC and will also be used in the detectors of
future electron-positron colliders. Multiple prototypes have been constructed
by the CALICE, CMS, and LHCb collaborations and have shown promising
performance in beam tests \cite{cms2017technical, Guz:2020mga, Boudry:2021quk}.
This concept proposes extremely high spatial segmentation. For instance, the
baseline ECAL of the CEPC has about three active cells per cubic centimeter.
Its longitudinal thickness of 24 radiation lengths ($\mathrm{X_0}$) is divided
into 30 sampling layers. Its transverse cell size is only $1 \times 1
  \rm{~cm^2}$. Such an ECAL can generate hundreds of hits, i.e., cells with a
signal above a readout threshold, for a 10 GeV photon or pion, as shown in Fig.
\ref{display}. The Particle Flow Algorithms (PFA) make full use of the HGC.
They combine the signals from the sub-detectors into a list of reconstructed
'particle flow objects', which ideally have a one-to-one correspondence with
the incident particles. The PFA includes a clustering algorithm and a matching
algorithm. The first one groups the calorimeter hits into clusters according to
the hit position, and the second module matches the clusters to the trajectory
in the tracker.

Further, HGC can be enhanced with the timing measurement of individual cells.
Depending on the applications, the precision can range from ns to 10's ps.
Several factors limit the precision. One natural factor is the signal
collection spread, typically the cell size divided by the speed of light.
Signal noise and clock jitter further degrade the performances at low and high
amplitudes, respectively \cite{Cavallari:2020xbu, apresyan2016investigation,
  akchurin2017timing}. This timing readout capability makes it possible to
measure a particle ToF by appropriately combining the cell information.
Averaging the times measured by multiple cells with charge-weighting, the
current silicon timing layers have shown time resolution higher than 25 ps
\cite{apresyan2016test, apresyan2016investigation, akchurin2018first}.
Moreover, measurement of the shower inner timing spectrum is hopeful of
extending the clustering of the PFA.

This work focus on the timing measurement using high-granularity ECAL's. After
a brief introduction of the involved detector configuration and simulation
software (Sec. \ref{sec2}), we analyse the shower true time spectrum and the
effect of intrinsic hit time resolution (Sec. \ref{sec3}). In Sec. \ref{sec4},
we propose a time reconstruction algorithm based on the quantile of the shower
time spectrum. We conclude that ToF resolutions of $5 \sim 20 \rm{~ps}$ for EM
showers and $80 \sim 160 \rm{~ps}$ for hadronic showers are achievable in the
CEPC ECAL. In section \ref{sec5}, we explore the dependence of time resolution
on: (a) the intrinsic time resolution of each individual channel and (b) the
number of timing readout layers. The expected timing performance of the CMS
CE-E is extrapolated from the result of the CEPC ECAL. It is affected by the
clustering algorithm, which is studied in Sec. \ref{sec6}. The next section
gives a brief summary. The algorithm proposed here is compared with an
alternative timing strategy in \ref{secA1}.

\section{Detector and software}\label{sec2}

\begin{figure}[htbp]
  \centering
  \includegraphics[width=0.43\textwidth]{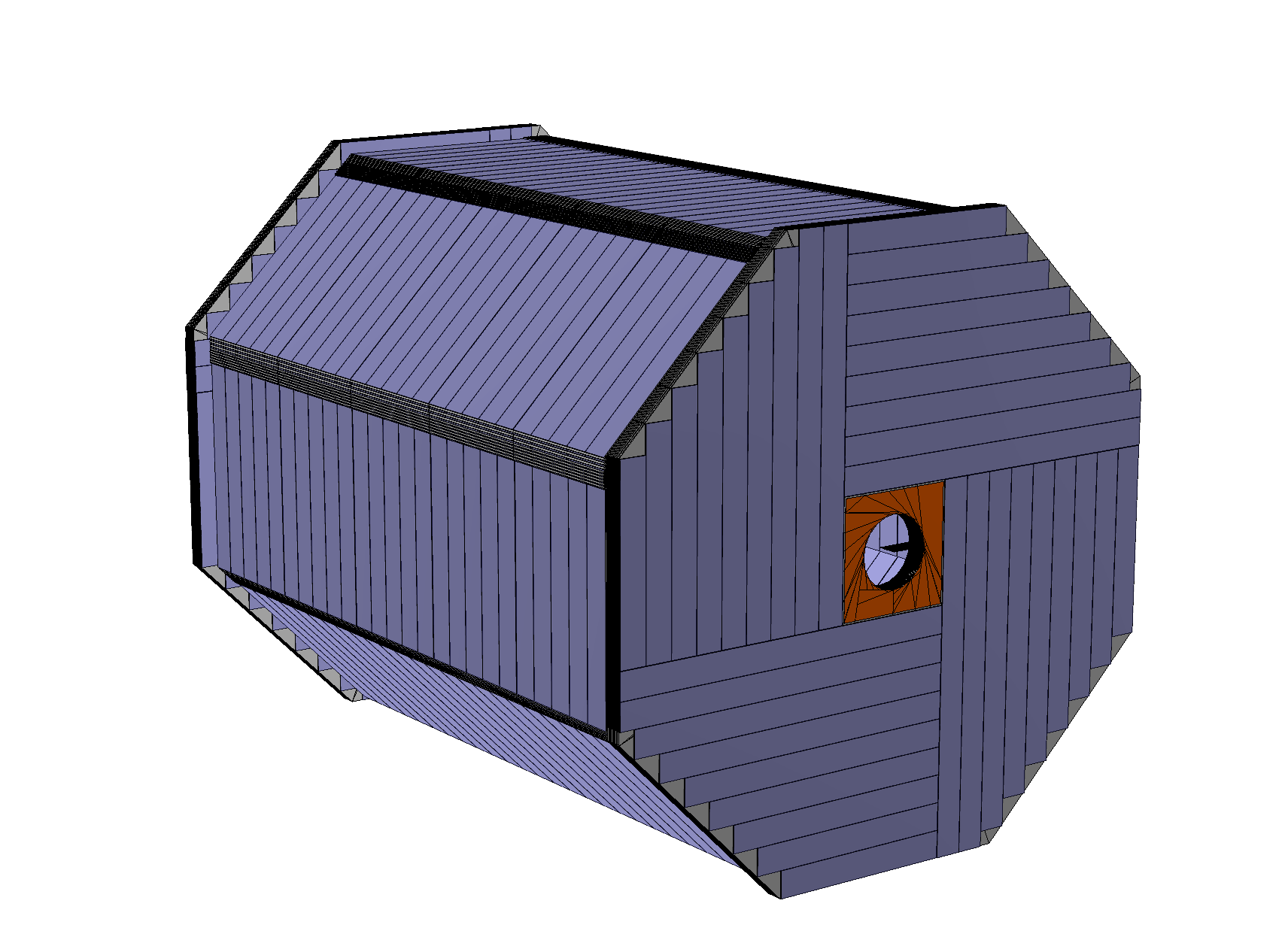}\\
  \includegraphics[width=0.43\textwidth]{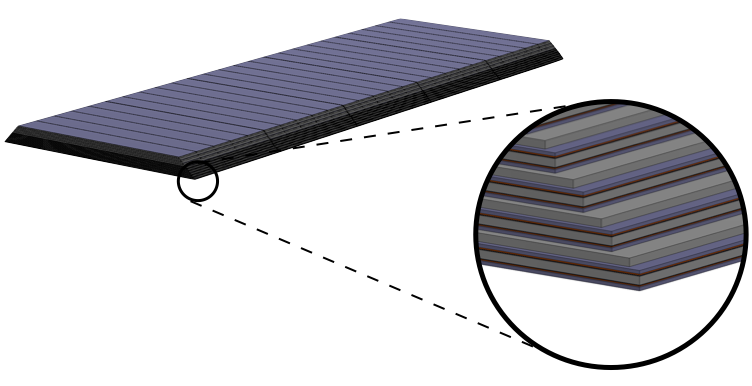}
  \caption{Geometry of the CEPC ECAL (top) and a display of one ECAL stave
    \cite{cepc2018cepcv2} (bottom).}
  \label{fig:ecal}
\end{figure}

\begin{figure}[htbp]
  \centering
  \includegraphics[width=0.5\textwidth]{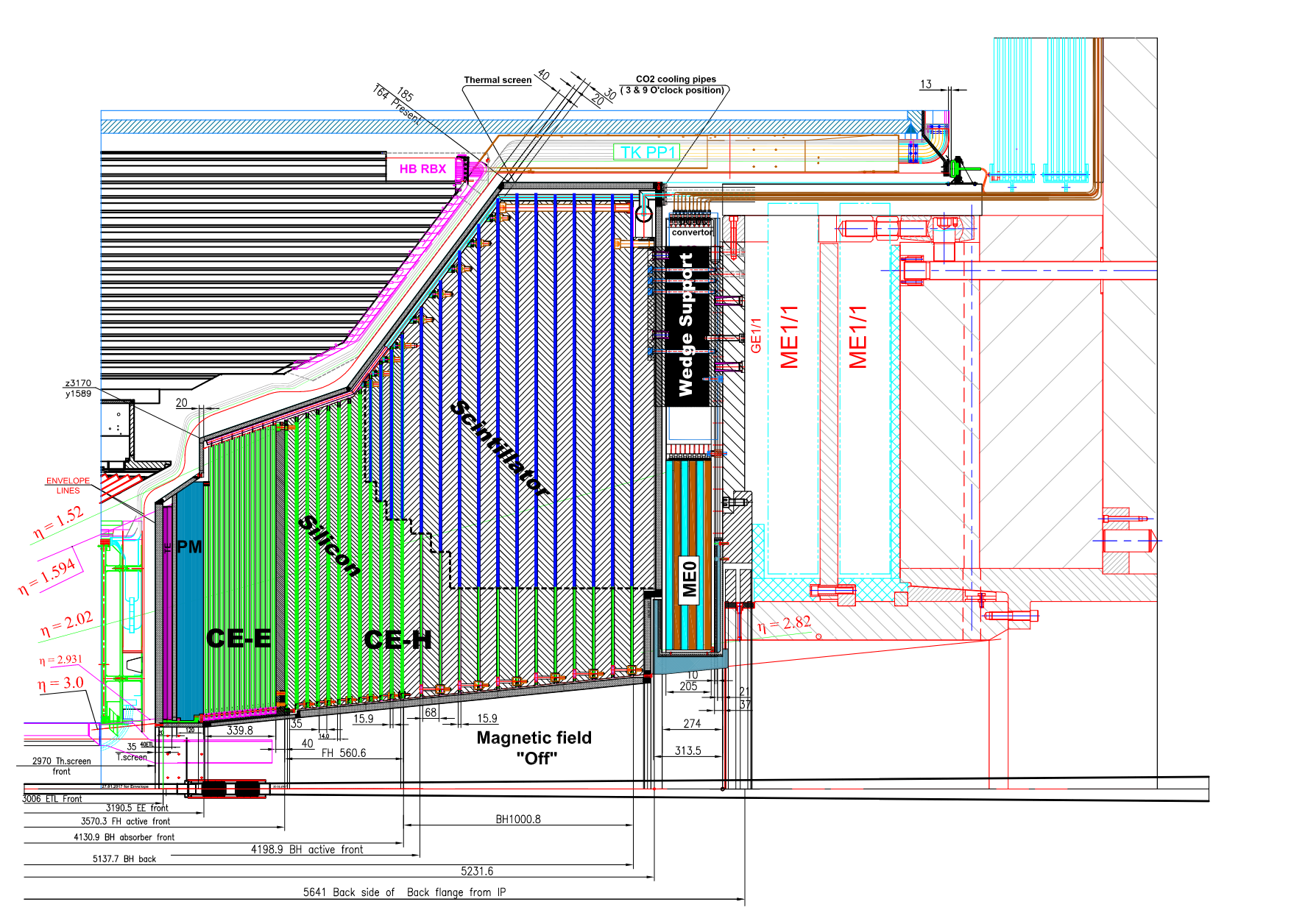}
  \caption{Longitudinal cross-section of the upper half of the CMS endcap
    calorimeter~\cite{cms2017technical}.}
  \label{fig:cms_hgcal}
\end{figure}

This study is based on the full simulation with the geometry of the CEPC
baseline ECAL. The ECAL geometry has been optimized based on the PFA
requirement. This calorimeter features an eight-stave barrel and two octagonal
endcap sections (Fig. \ref{fig:ecal}). The inner radius of the barrel is 1847
mm, and the distance between the Interaction Point (IP) and the front face of
the ECAL endcap is 2450 mm. In the radial direction, the ECAL is segmented into
30 sampling layers, each consisting of a tungsten absorber and an active layer.
The thickness of the tungsten plates in the first 20 layers is 0.6 radiation
lengths ($\mathrm{X_0}$), and double in the last ten layers. Each active layer
is equipped with square matrices of highly resistive silicon diodes, segmented
in cells of $10 \times 10 \rm{~mm^2}$ and a thickness of 0.5 mm. The CEPC ECAL
includes around $2\times10^{7}$ channels.

The HGC concept is also applied to upgrading the CMS endcap calorimeters, now
in construction, whose geometry is shown in Fig. \ref{fig:cms_hgcal}. It
comprises an electromagnetic compartment (CE-E) and a hadronic calorimeter
(CE-H). The CE-E is a sampling calorimeter equipped with silicon sensors and
tungsten absorbers. Along the direction of the beam pipe, the CE-E is installed
in the range from $\left|z\right| = 3.19 \rm{~m}$ to $3.53 \rm{~m}$, for an IP
at $z = 0$. In the cross-section perpendicular to the beam pipe, CE-E has a
disk-like shape with an inner and outer radius of 0.32 m and 1.68 m at the
front face. The total longitudinal thickness of $27.7~\mathrm{X_0}$ is divided
into 26 sampling layers \cite{Lange:2022gc}. The radiation fluence increases
with the increasing pseudorapidity ($\eta$). In order to maintain the
performance after the integrated luminosity of $3000 \rm{~fb^{-1}}$, the
disk-like CE-E is divided into three rings corresponding to $r = 35 \sim 70
  \rm{~mm}$, $70 \sim 100 \rm{~mm}$ and $100 \sim 180 \rm{~mm}$. From the inside
out, the radiation fluence decreases. Three types of silicon sensors with
deployment thickness of $120\rm{~\mu m} \times 0.52 \rm{~cm^2}$, $200\rm{~\mu
    m} \times 0.52 \rm{~cm^2}$, $300\rm{~\mu m} \times 1.18 \rm{~cm^2}$ are
equipped in these three regions. The total number of channels is
$3.916\times10^{6}$. The granularity information of the CE-E and the CEPC ECAL
is shown in Tab. \ref{tab:geometry}. In this paper, we extrapolate the
evaluated timing performance in the CEPC ECAL to that in the CMS CE-E.

To quantify the ToF performance of the CEPC ECAL, we simulate ECAL showers
resulting from single $\Pphoton$, $\Pelectron$, $\Ppiplus$, $\PKplus$ and
$\Pproton$ using a full simulation package based on
GEANT4~\cite{agostinelli2003geant4}. For reference, we also simulated the
energy deposition of muons, which are close to minimum ionization particles
(MIP). The momentum of the created particles is at a single value or uniformly
distributed in a range from 0 to 30 GeV. The particle originates from a point
193 mm from the IP\footnote{To avoid early interactions} and is shot
perpendicular to the ECAL surface. The magnetic field is turned off so that
charged particles propagate to ECAL along straight lines, and the redundant
discussion about trajectory correction is simplified. The statistic of the
involved samples is listed in Tab. \ref{tab:sample}.

\begin{table}[htbp]
  \centering
  \caption{Geometry, granularity and active material of the CMS HGCAL(CE-E)
    \cite{cms2017technical,Lange:2022gc} and the CEPC ECAL~\cite{cepc2018cepcv2}.}
  \label{tab:geometry}
  \begin{tabular}{ccc}
    \hline
                       & CMS HGCAL                   & CEPC ECAL         \\ \hline
    Sensitive material & Silicon                     & Silicon           \\
    Layers number      & 26                          & 30                \\
    Total thickness    & 27.7 $\mathrm{X_0}$         & 24 $\mathrm{X_0}$ \\
    Cell size          & $\sim 0.5/1.1\mathrm{cm}^2$ & 1 $\mathrm{cm}^2$ \\
    Cell shape         & hexagon                     & square            \\ \hline
  \end{tabular}
\end{table}

\begin{table}[htbp]
  \centering
  \caption{Statistics of the simulated single particle samples. Six samples were generated
    with uniformly distributed incident momentum, while three samples with single incident
    momentum.}
  \label{tab:sample}
  \scalebox{0.75}{
    \begin{tabular}{ccccccc}
      \hline\noalign{\smallskip}
      Particle     & $\Pphoton$       & $\Pelectron$     & $\Pmuon$         & $\Ppiplus$         & $\PKplus$          & $\Pproton$         \\
      \noalign{\smallskip}\hline\noalign{\smallskip}
      Energy [GeV] & $0 - 30$         & $0 - 30$         & $0 - 30$         & $0 - 30$           & $0 - 30$           & $0 - 30$           \\
      Event Num.   & $3\times 10^{4}$ & $3\times 10^{4}$ & $3\times 10^{4}$ & $3.6\times 10^{4}$ & $3.6\times 10^{4}$ & $3.6\times 10^{4}$ \\
      \noalign{\smallskip}\hline\noalign{\smallskip}
      Energy [GeV] & $10$             & -                & $10$             & $10$               & -                  & -                  \\
      Event Num.   & $5\times 10^{3}$ & -                & $5\times 10^{3}$ & $5\times 10^{3}$   & -                  & -                  \\
      \noalign{\smallskip}\hline
    \end{tabular}
  }
\end{table}

\section{Shower energy-time spectrum}\label{sec3}
\begin{figure*}[t]
  \centering
  \begin{minipage}{0.28\textwidth}
    \caption{The distribution of the true projected hit time in the range of $< 6 \mathrm{~ps}$ in the 10 GeV $\Pphoton$, $\Ppiplus$ and $\Pmuon$ samples (left) and the corresponding cumulative distribution in the time range of $0\sim 1 \rm{~\mu s}$ (right). The dashed lines in the left plot are the expected time when the incident particles reach the front of the ECAL. In the right plot, the solid black line corresponds to the boundary (6 ps) of the left plot. To show the complete cumulative distribution, the x-axis of the right plot uses a symmetrical logarithmic scale, which is linear in the range of $\left[-2, 2\right] \rm{~ps}$ and logarithmic in the other region. The total number of all the hits earlier than $1 \rm{~\mu s}$ is normalized to unity in these two plots, which means the integral of the distribution in the left plot equals to the corresponding value in the right cumulative distribution.}

    \label{g_hit_truth_delay}
  \end{minipage}
  \begin{minipage}{0.7\textwidth}
    \centering
    \includegraphics[width=0.47\textwidth]{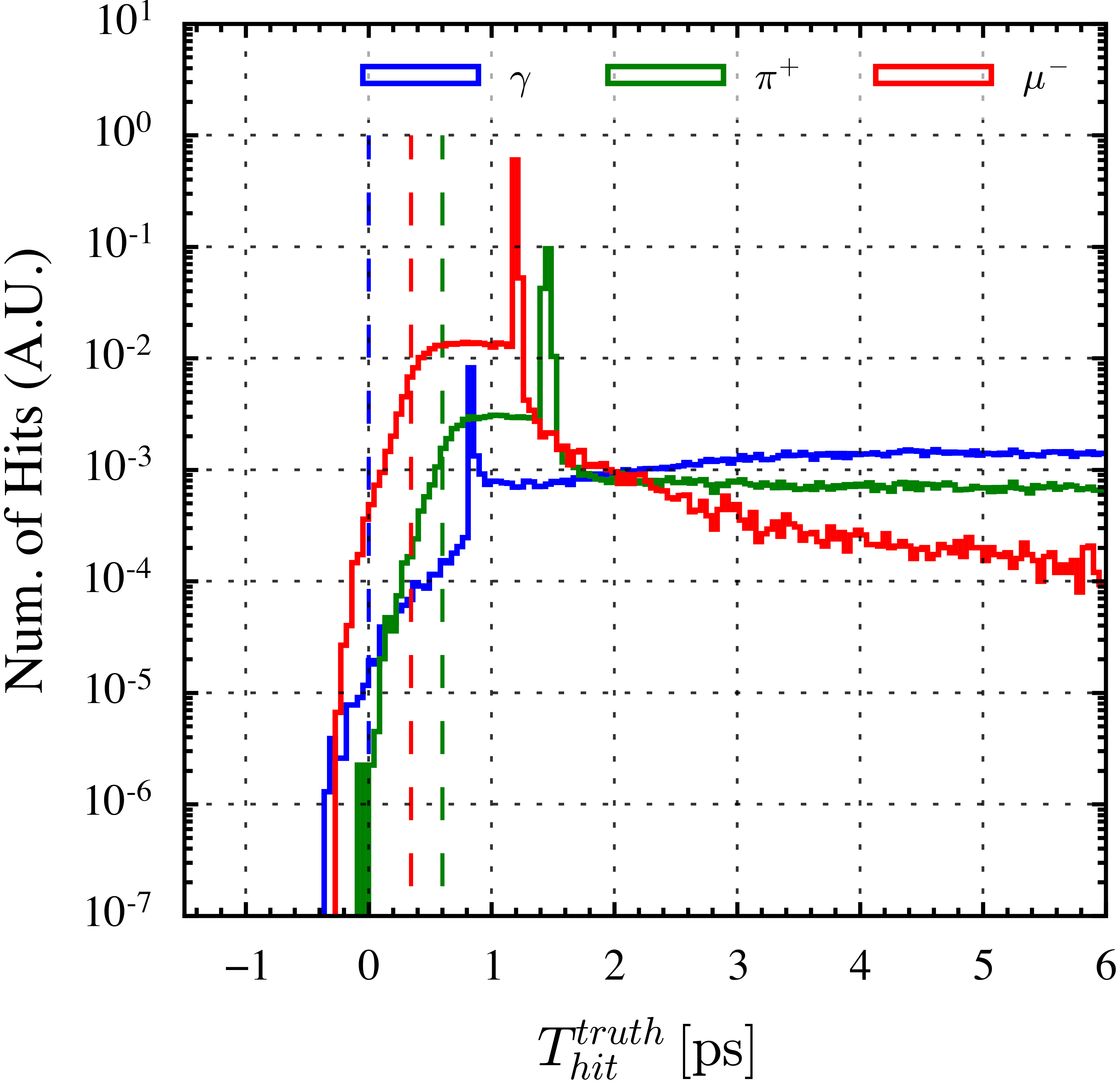}\quad
    \includegraphics[width=0.475\textwidth]{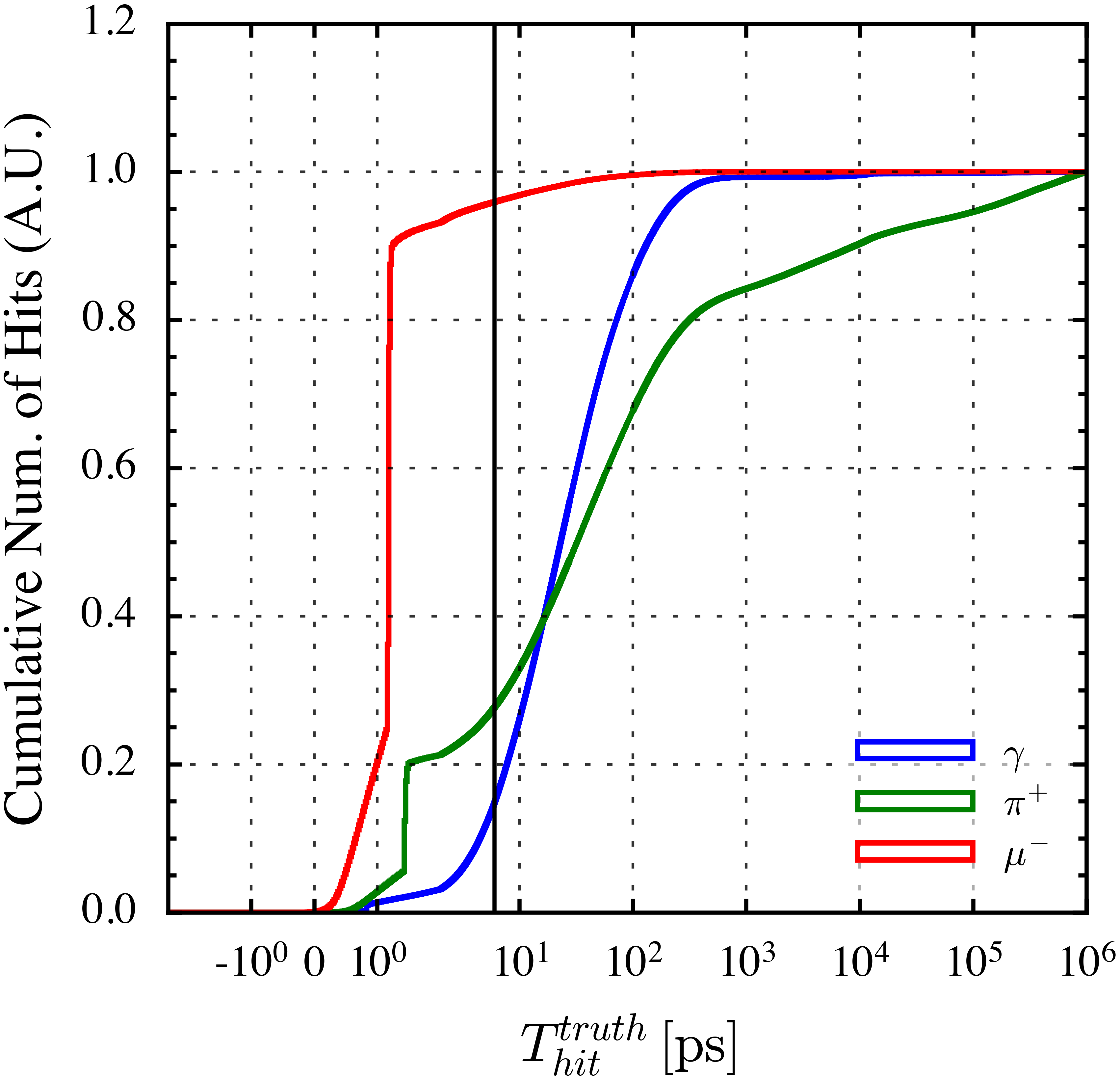}
  \end{minipage}
\end{figure*}

\begin{figure*}[ht]
  \centering
  \includegraphics[width=0.32\textwidth]{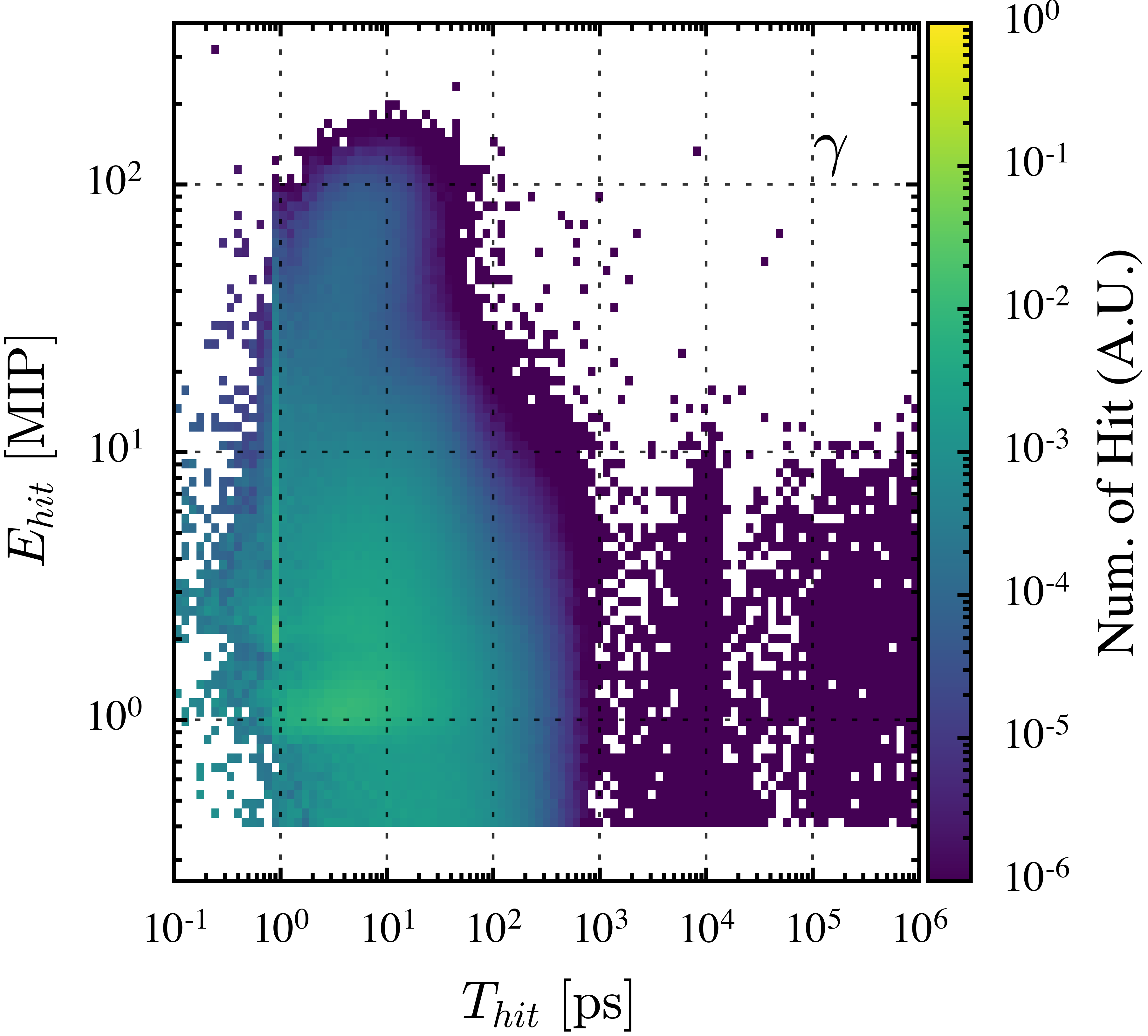}\quad
  \includegraphics[width=0.32\textwidth]{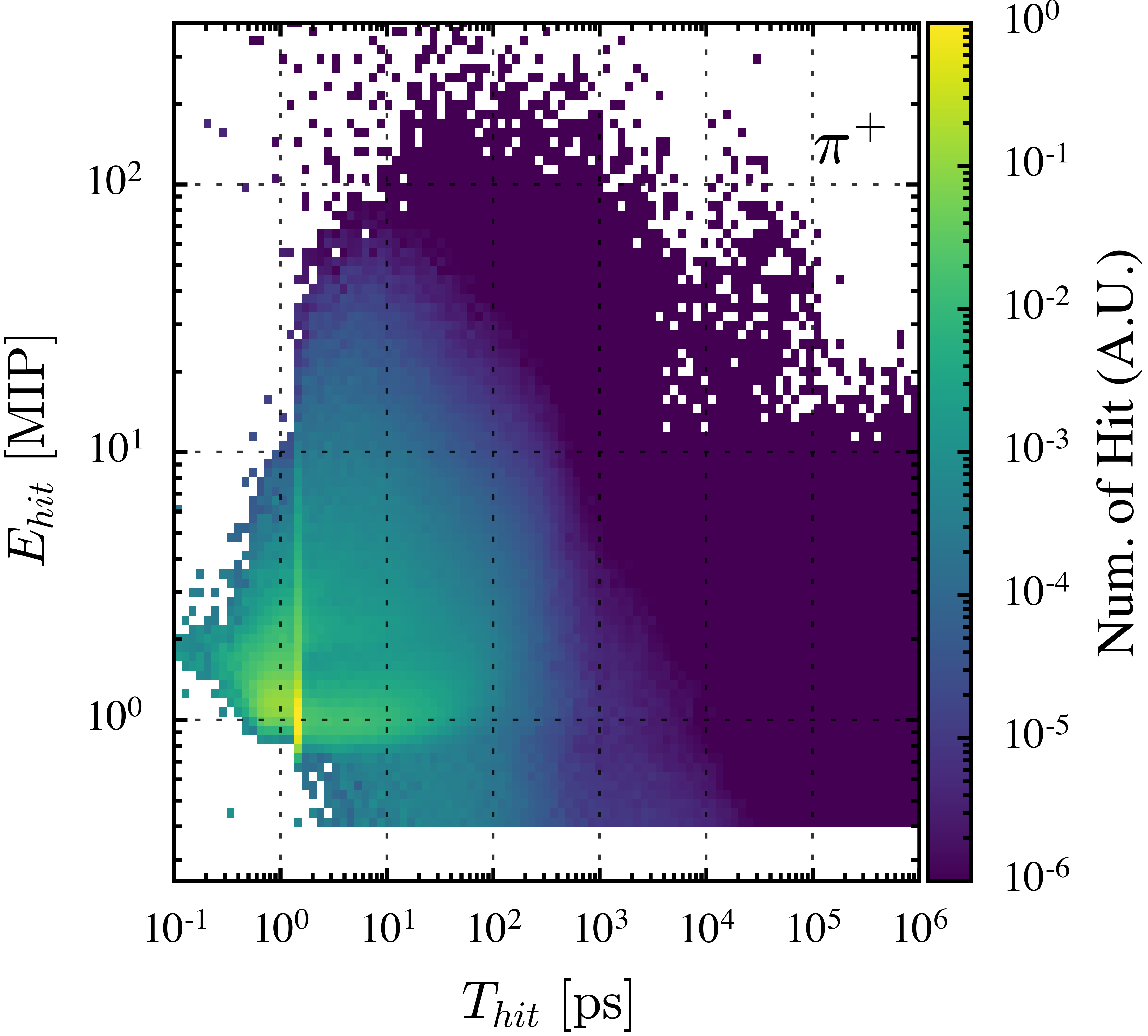}\quad
  \includegraphics[width=0.32\textwidth]{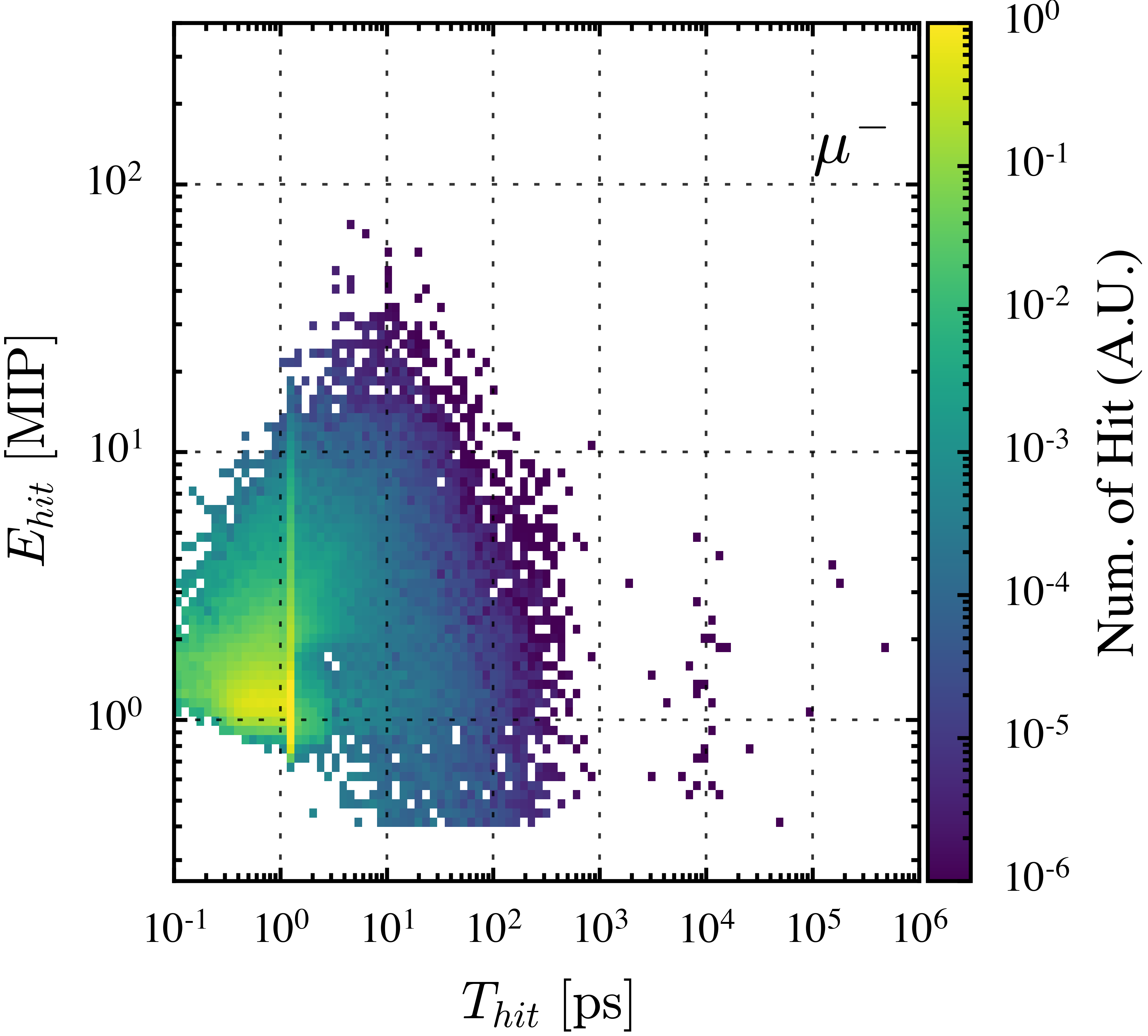}\quad
  \caption{The two-dimensional probability density distribution of the hits projected time and energy in 10 GeV $\Pphoton$ (left), $\Ppiplus$ (middle) and $\Pmuon$ (right) samples. The total number of all the hits earlier than $1 \rm{~\mu s}$ is normalized to unity.}
  \label{g_en_delay}
\end{figure*}

We first investigate the distribution of the hit time at the truth level.
Secondly, we implement a time digitization process according to the CMS beam
test result on the intrinsic time resolution of the single silicon sensor
\cite{akchurin2017timing}. We discuss the digitized time spectrum pattern,
which is supposed to be fully accessible at the CEPC ECAL.

The true information of each hit is extracted from the steps and tracks in the
cell generated by Geant4. The zero time is fixed to when the particle is
created. The energy of the hits given in GeV is normalized in units of MIPs
(0.147 MeV), where a MIP is defined as the most probable value of the energy
deposition in a silicon sensor by a 10 GeV muon hitting perpendicularly. Only
the hits with energy higher than 0.05 MeV or about 1/3 MIPs are considered in
our analysis. Hits occurring after $1 \mathrm{~\mu s}$ are ignored, this time
threshold is larger than the time spread of pile-up events at the HL-LHC and
the time spacing of the bunch crosses at the CEPC. To perfectly reflect the
behavior of the timing electronics, a good definition of the true times of cell
hits is to select the earliest time above an energy threshold. However, the
energy threshold depends strongly on the specific discriminator in the
electronic. In this study, true hit times are defined as the time of the most
energetic step in the cell. This approximation differs only slightly from the
previous definition because the cases of multiple energetic but well-separated
depositions in time are rare.

Fig. \ref{g_hit_truth_delay} shows the true hit time spectrum of 10 GeV
photons, charged pions, and muons, as well as the expected ToFs of these three
types of particles. In these plots, the time is redefined as the projected
time,
\begin{equation}\label{normalization}
  T = t - L/c
\end{equation}
where $t$ denotes the raw hit times, $L$ denotes the distance between the IP and the center of the hit. Since the magnetic field is set to zero, this subtraction approximately normalizes the propagation time from the IP to the ECAL hit. Each distribution contains a fast component from 0 to 2 ps, followed by a slow tail extending beyond one ns. During the simulation, there are a large amount of hits including only one step and pausing a peak in the fast component. The hits with multiple steps contribute to the platform before the peak. Moreover, the finite granularity of the ECAL causes an error in $L$. Consequently, there is a small fraction of shower hits whose projected time is before zero.

The three plots in Fig. \ref{g_en_delay} show the correlation between the hit
time and energy for showers produced by 10 GeV $\Pphoton$'s, $\Ppiplus$'s and
$\Pmuon$'s. A large fraction of hits has energy lower than 5 MIPs. The energy
deposition corresponding to one and two MIPs are visible. The shadows around 10
ns for photons arises from the back-scattering hits in the opposite side of the
ECAL. In addition, the photon and pion showers include an extremely slow
component later than 100 ps with a faction of 15\% and 30\%, respectively,
which may constitute noise in the next physics event. More importantly for the
timing of the showers, the energetic hits tend to occur in the fast component
of the showers, especially in the EM showers. The energy of the hits with
projected time from 0 to 100 ps ranges from several MIPs to around 100 MIPs in
the first plot of Fig. \ref{g_en_delay}.

\begin{figure}[htbp]
  \centering
  \includegraphics[height=0.43\textwidth]{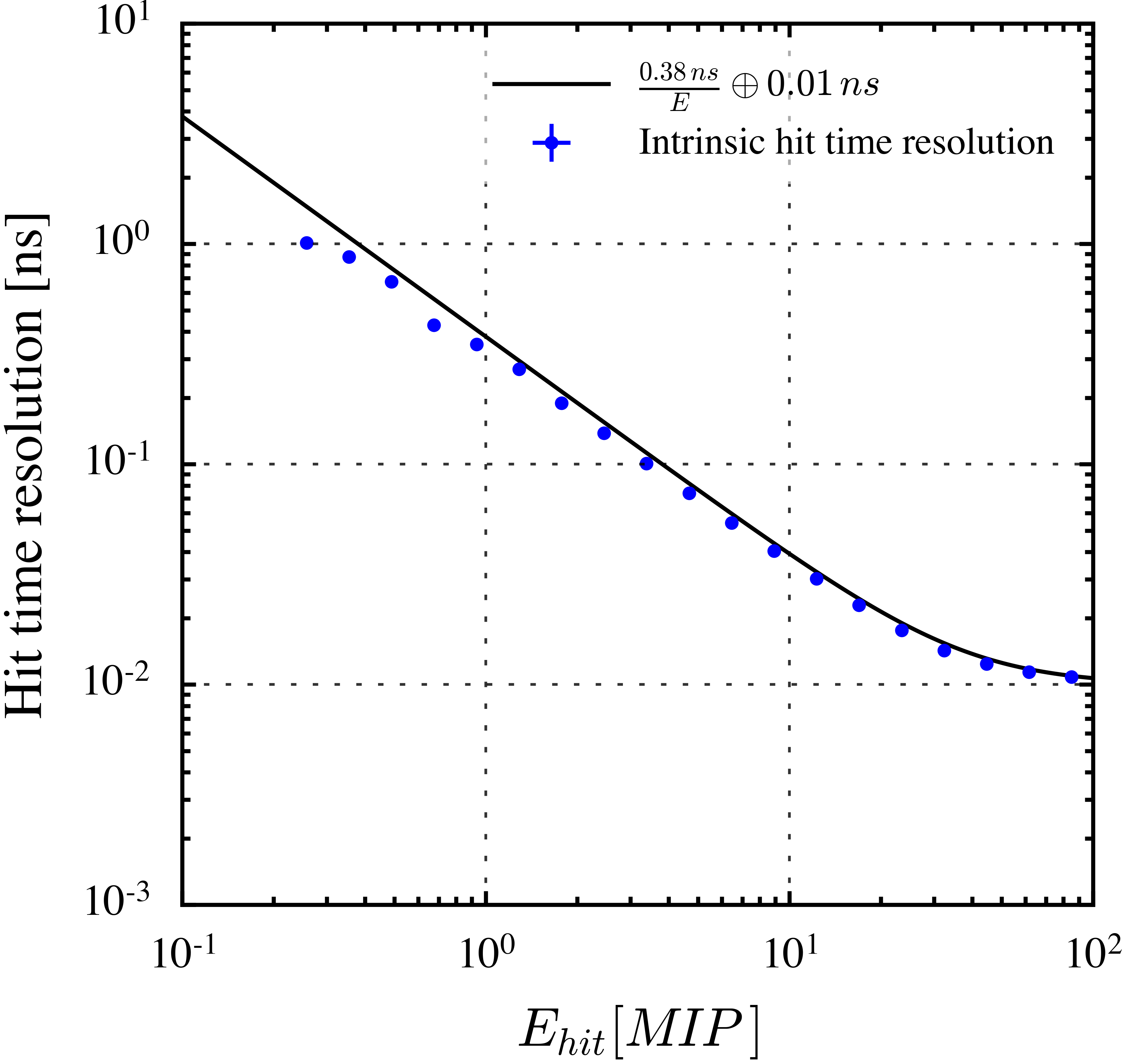}
  \caption{The intrinsic hit time resolution as a function of the energy deposition in the silicon sensors. The black solid line is the model of CMS measurement \cite{akchurin2017timing} and the blue dots are the result of the digitization in the 10 GeV $\Pmuon$ samples.}
  \label{g_int_hit_reso}
\end{figure}

To mimic the effect of the detector and its readout electronics, we implement a
digitization process based on the CMS beam test results on the time response of
the thin planar silicon diodes \cite{akchurin2017timing}. In the CMS report,
the intrinsic hit time resolution has been measured by the independent returns
of two parallel sensors, as

\begin{equation}
  \frac{\sigma(t_{1} - t_{2})}{\sqrt{2}} = \frac{A}{\sqrt{2}S_{\mathrm{eff}}} \oplus C,
  \label{eq:cms_dt}
\end{equation}

\noindent where $S_{\mathrm{eff}} = S_{1}S_{2}/\sqrt{S_{1}^{2} + S_{2}^{2}}$ and $S_{1}$, $S_{2}$ denotes the signal strength of the two sensors. For the sensors with depletion thickness of $211\mathrm{\mu m}$, the coefficients A and C are respectively $380\;\mathrm{ps \cdot MIP}$ and $10 \mathrm{ps}$. Accordingly, we smeared the simulated true time of each hit with a Gaussian response function. The width of the response function equals the intrinsic hit time resolution parameterized as $\sigma = \frac{A}{E} \oplus C, ~A = 380\;\mathrm{ps \cdot MIP} \rm{~and~} C = 10 \mathrm{ps}$. $E$ denotes the energy deposition in MIP units and is equivalent to $\sqrt{2} S_{\mathrm{eff}}$. The resulting time resolution as a function of hit energy is shown in Fig. \ref{g_int_hit_reso}. When the hit energy is higher than 100 MIPs, the intrinsic hit time resolution saturates at 10 ps. However, a large amount of hits in the showers have only an energy of several MIPs, for which the intrinsic hit time resolution is worse than 100 ps.

\begin{figure}[htbp]
  \centering
  \includegraphics[height=0.41\textwidth]{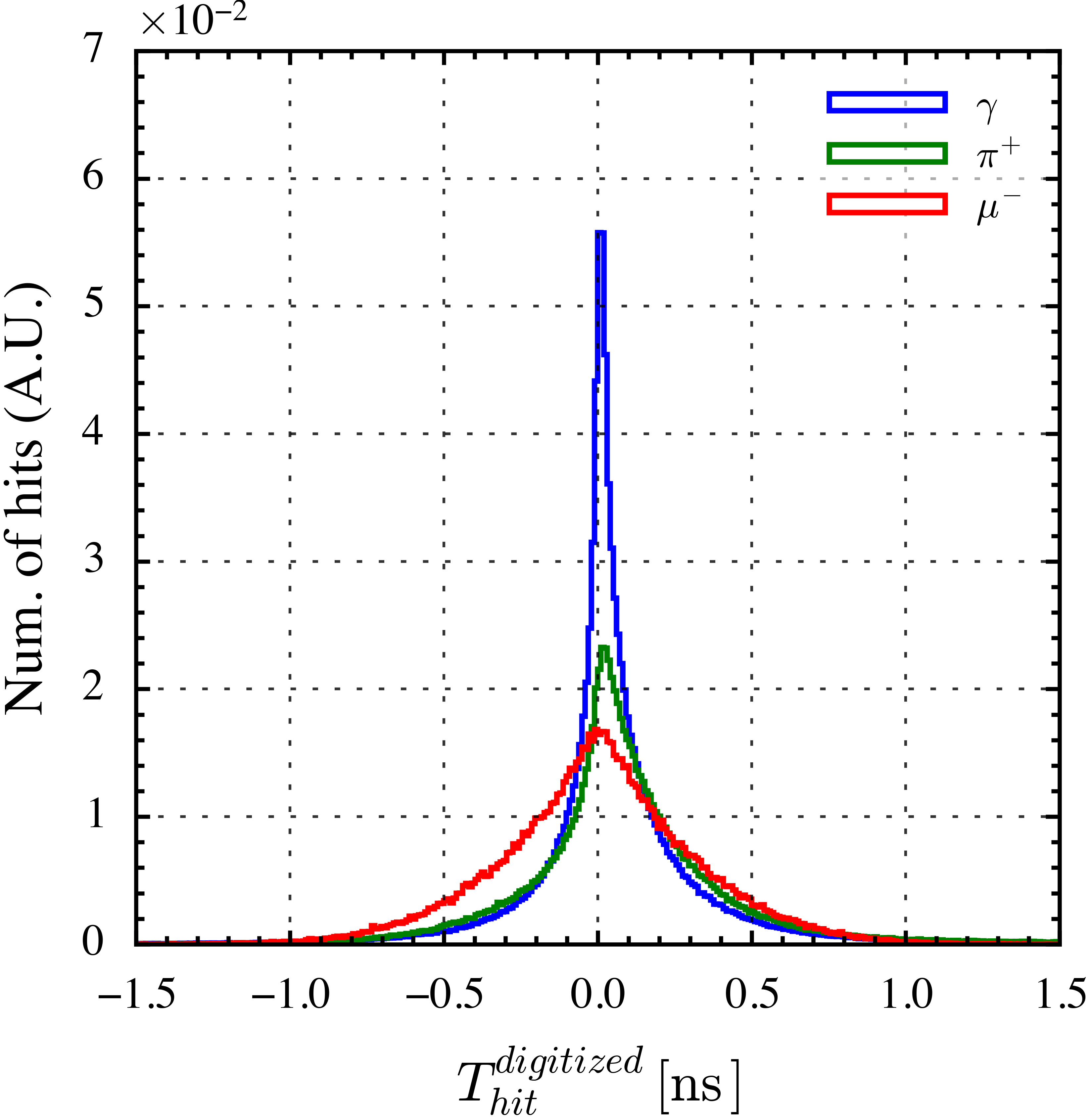}
  \quad
  \includegraphics[height=0.41\textwidth]{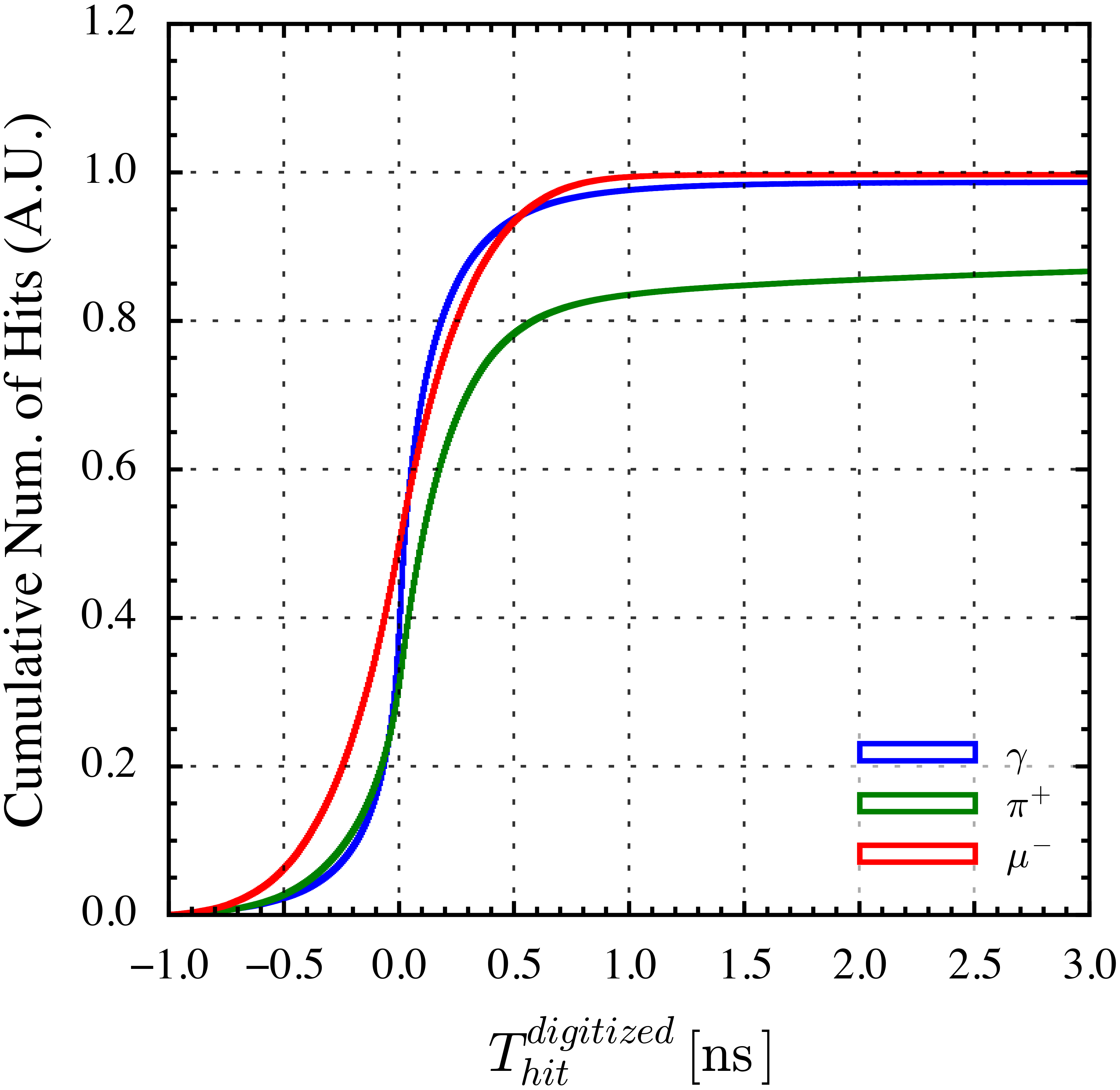}
  \caption{The digitized hit time distribution in range of $-1\sim 1.5 \mathrm{~ns}$ of 10 GeV $\Pphoton$, $\Ppiplus$ and $\Pmuon$ (top) and the cumulative distribution in the time range of $-1\sim 3$ ns (bottom), where the total number of all the hits earlier than 1000 ns is normalized to unity in these two plots..}
  \label{g_hit_digi_delay}
\end{figure}

Fig. \ref{g_hit_digi_delay} shows the projected time spectrum of showers after
time digitization. The fast component, which occurs before the first ten ps on
the truth level (Fig. \ref{g_hit_truth_delay}), is smeared into the $-0.5$ to
$0.5$ ns region, where the digitization uncertainty dominates the shape. The
shape of the fast component distribution is a highly non-Gaussian peak, as it
is a combination of various resolutions. The slow component at the truth level
is also retained after digitization, and its fraction differs according to the
shower types. The fraction of the hits with projected time above one ns is
almost zero in photon and muon but about 20\% in pion showers.

\section{A ToF reconstruction algorithm}\label{sec4}

Considering the non-Gaussian distribution and the later tail of the digitized
shower time, a blind average of the hits brings many biases. We propose a
universal shower time reconstruction algorithm based on a given quantile of the
projected hit times.

Starting with the collection of all digitized hit times of the cluster, we sort
the times in ascending order and use the value of the $(R\cdot
  N_{\rm{hits}})$-th as a result, where $R$ is an ad hoc ratio and
$N_{\rm{hits}}$ is the number of the shower hits. When $R < 0.5$, the result
corresponds to the median of the fastest $2R\cdot N_{\rm{hits}}$ hit time.

The single parameter $R$ should be optimized for a reasonable time
reconstruction. In the following section, we qualify the algorithm's
performance in bias and resolution and discuss their behavior under variations
of $R$.

\subsection{Performance for single particle showers}\label{sec4.1}

\begin{figure*}[t]
  \centering
  \begin{minipage}[t]{0.28\textwidth}
    \vspace{0pt}
    \caption{The distribution of the difference between reconstructed shower time and the true time in the 10 GeV $\Ppiplus$ sample. To remove the outliers, a time residual window (red lines) is defined as $\left[Q_2 - 5(Q_3 - Q_1), Q_2 + 5(Q_3 - Q_1)\right]$, where $Q_1$, $Q_2$ and $Q_3$ are the three quartiles of the distribution. The bias and resolution is defined as the mean and standard deviation of the data inside the window.}
    \label{fig:residual}
  \end{minipage}
  \begin{minipage}[t]{0.7\textwidth}
    \vspace{0pt}
    \centering
    \includegraphics[width = 0.7\textwidth]{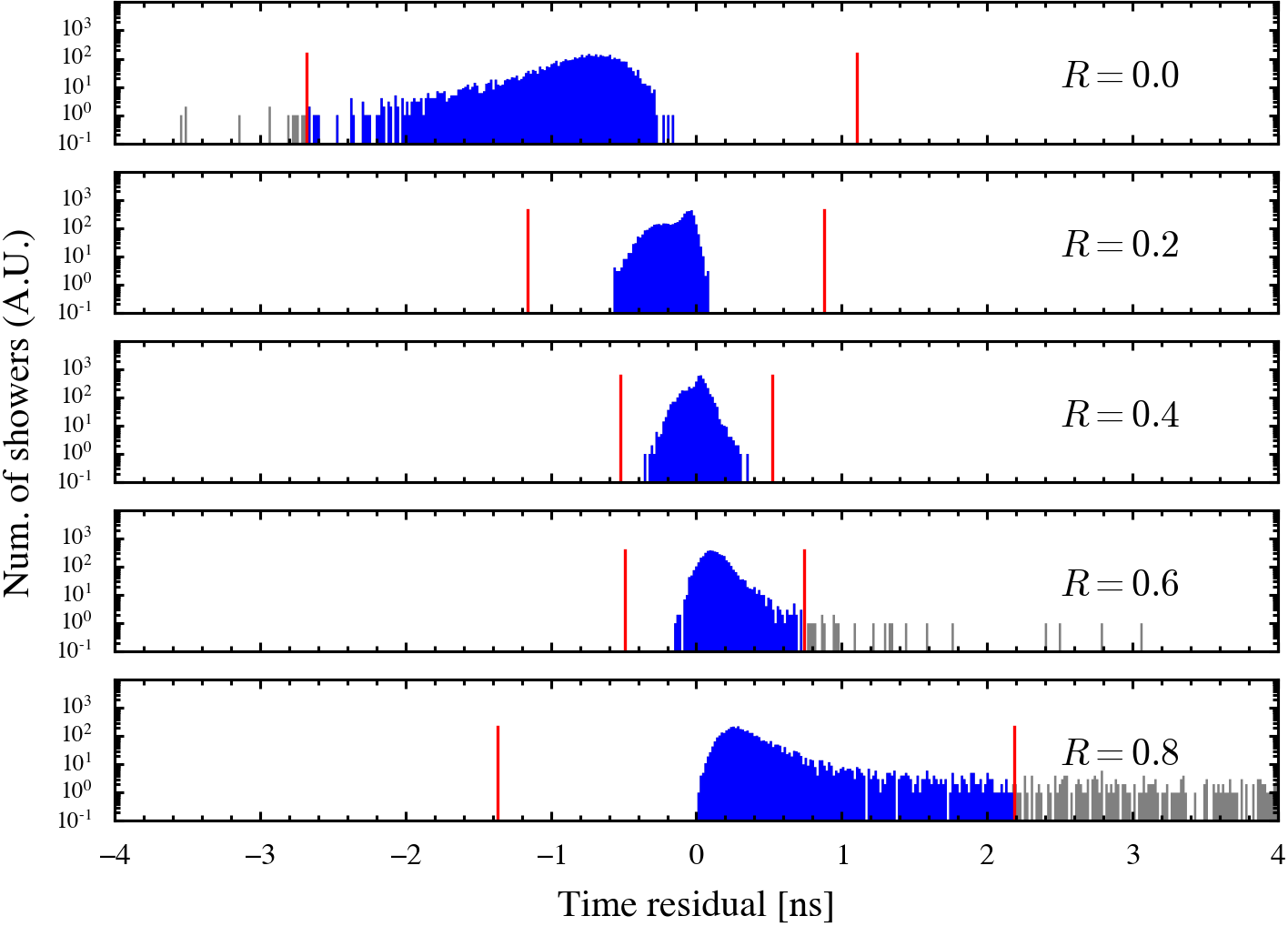}
  \end{minipage}
\end{figure*}

\begin{figure*}[ht]
  \centering
  {\includegraphics[width=0.32\textwidth]{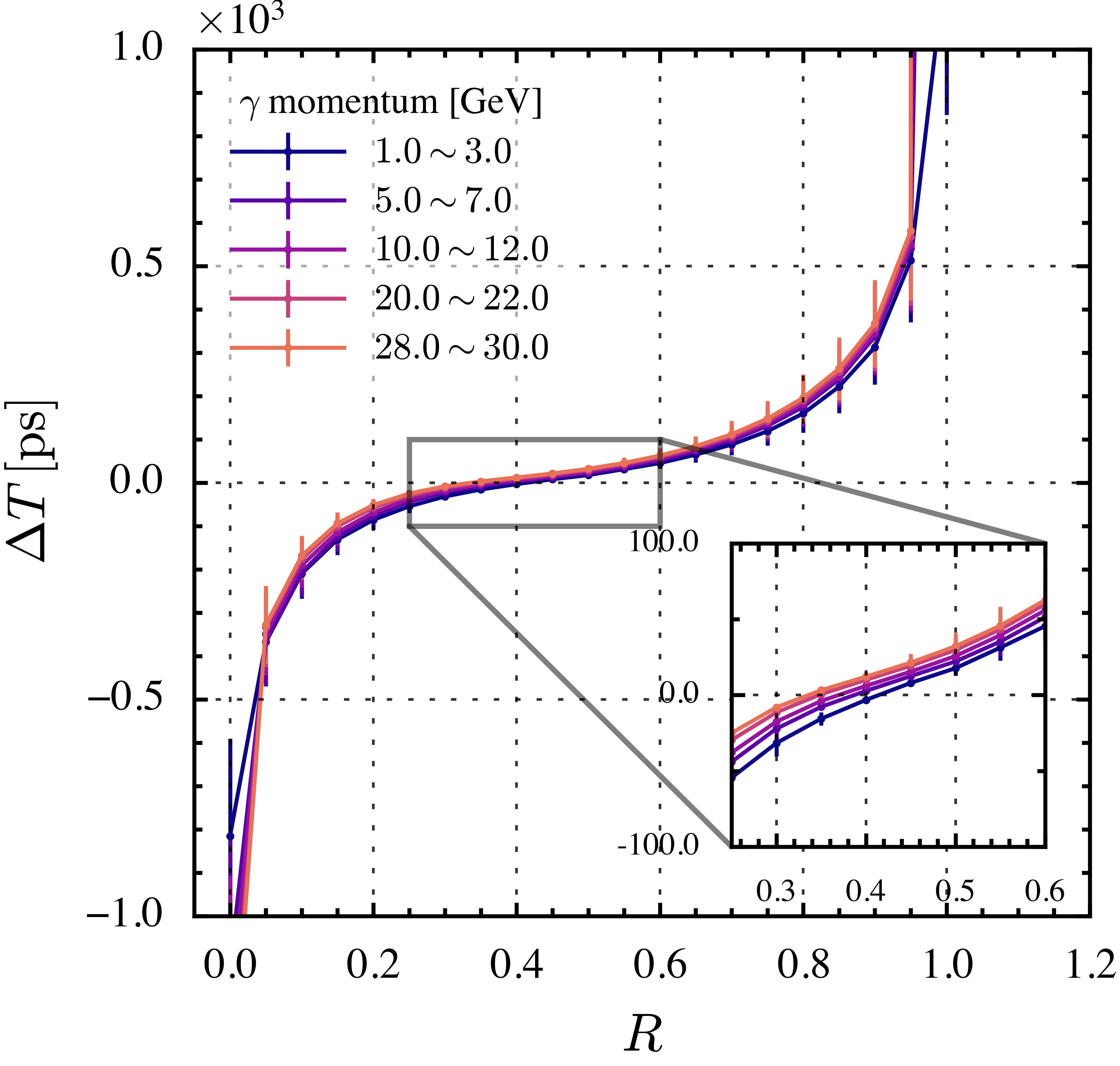}}	
  {\includegraphics[width=0.32\textwidth]{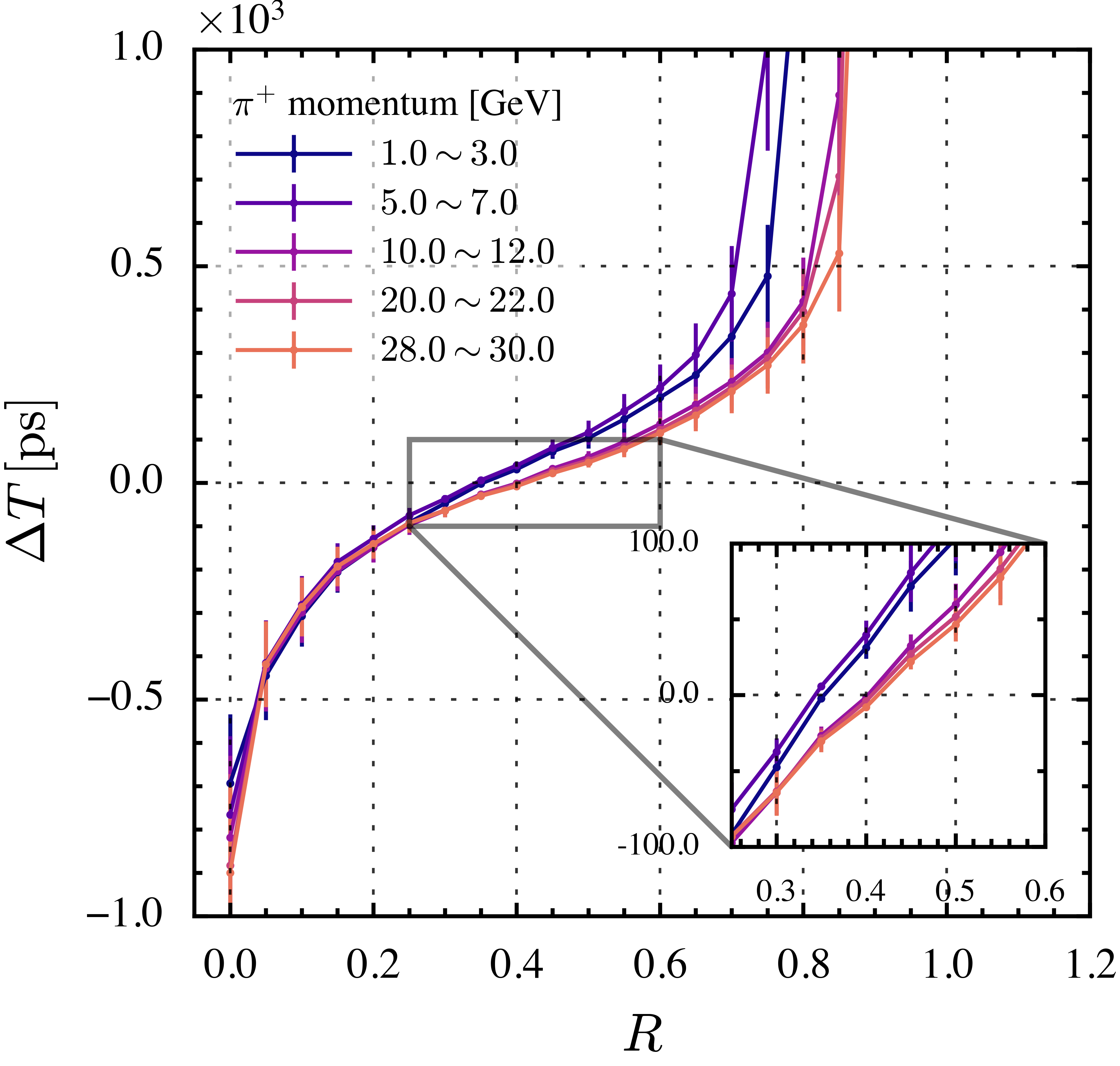}} 
  {\includegraphics[width=0.32\textwidth]{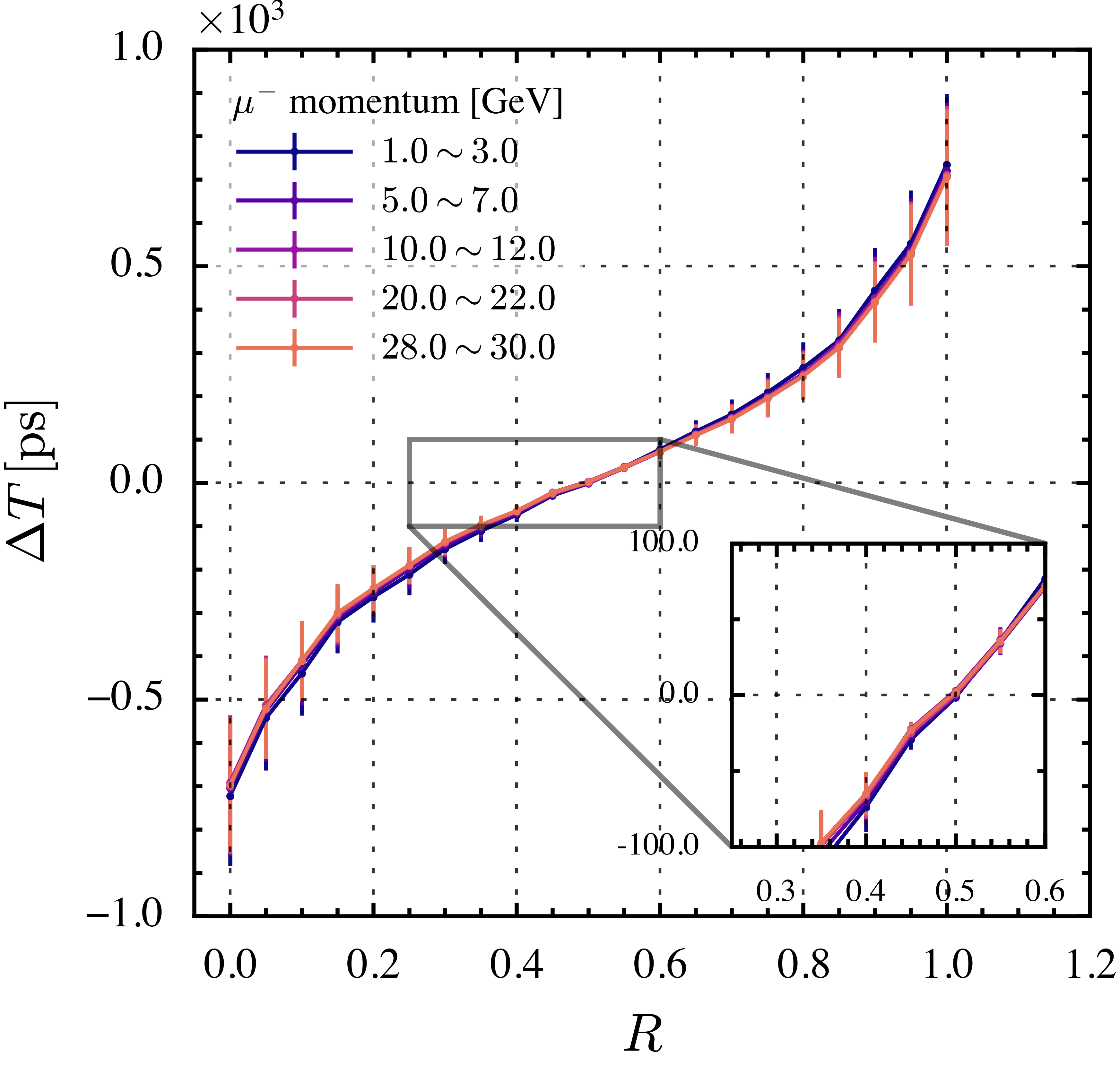}}

  \caption{Bias of reconstructed shower time in $\Pphoton$ (left), $\Ppiplus$ (middle), $\Pmuon$ (right) samples as a function of $R$. The errors are all scaled with a factor of ten, and also in Fig. \ref{reso_curve}.}
  \label{bias_curve}
\end{figure*}

\begin{figure*}[ht]
  \centering
  \subfigure
  {\includegraphics[width=0.32\textwidth]{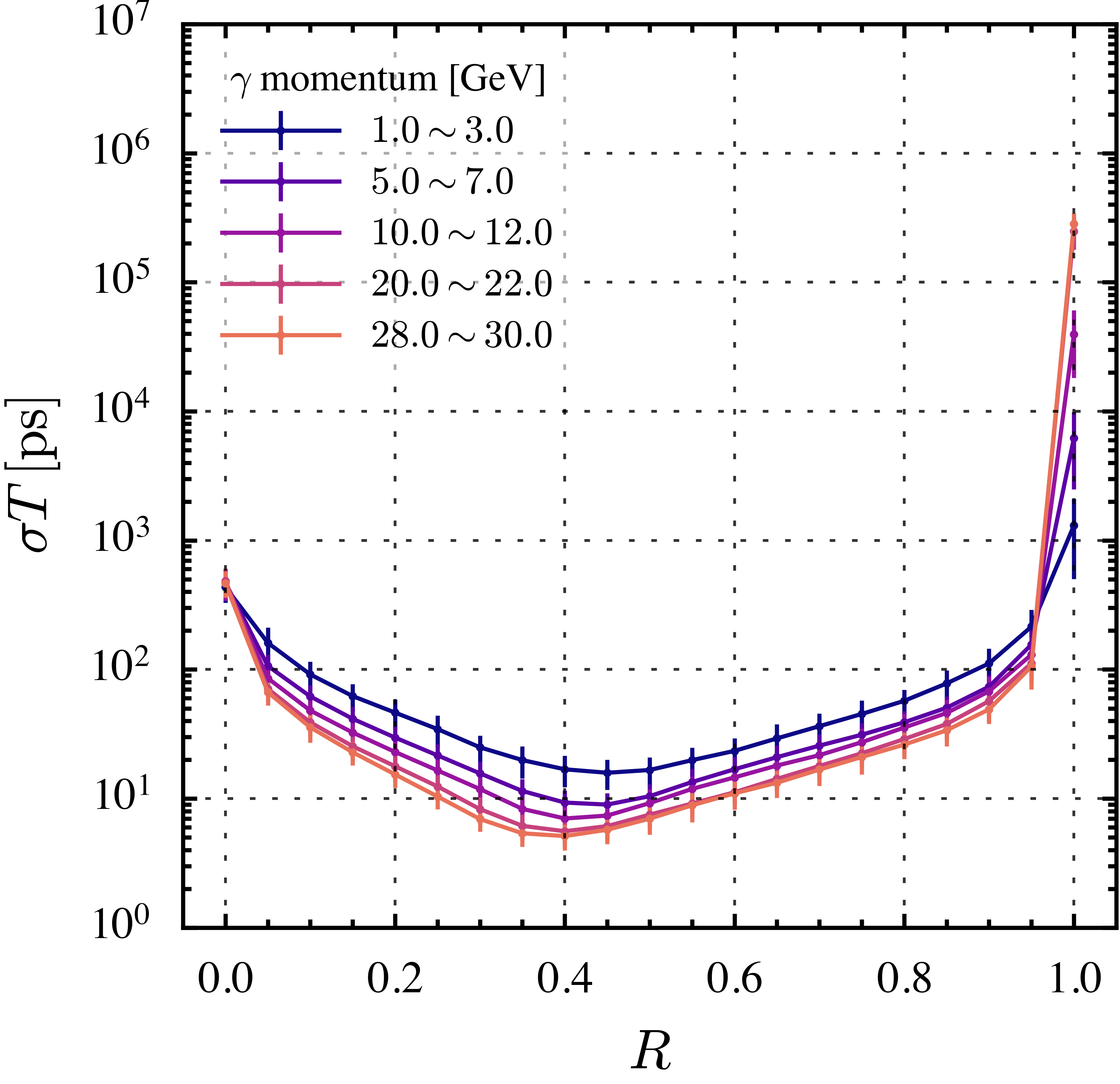}} {\includegraphics[width=0.32\textwidth]{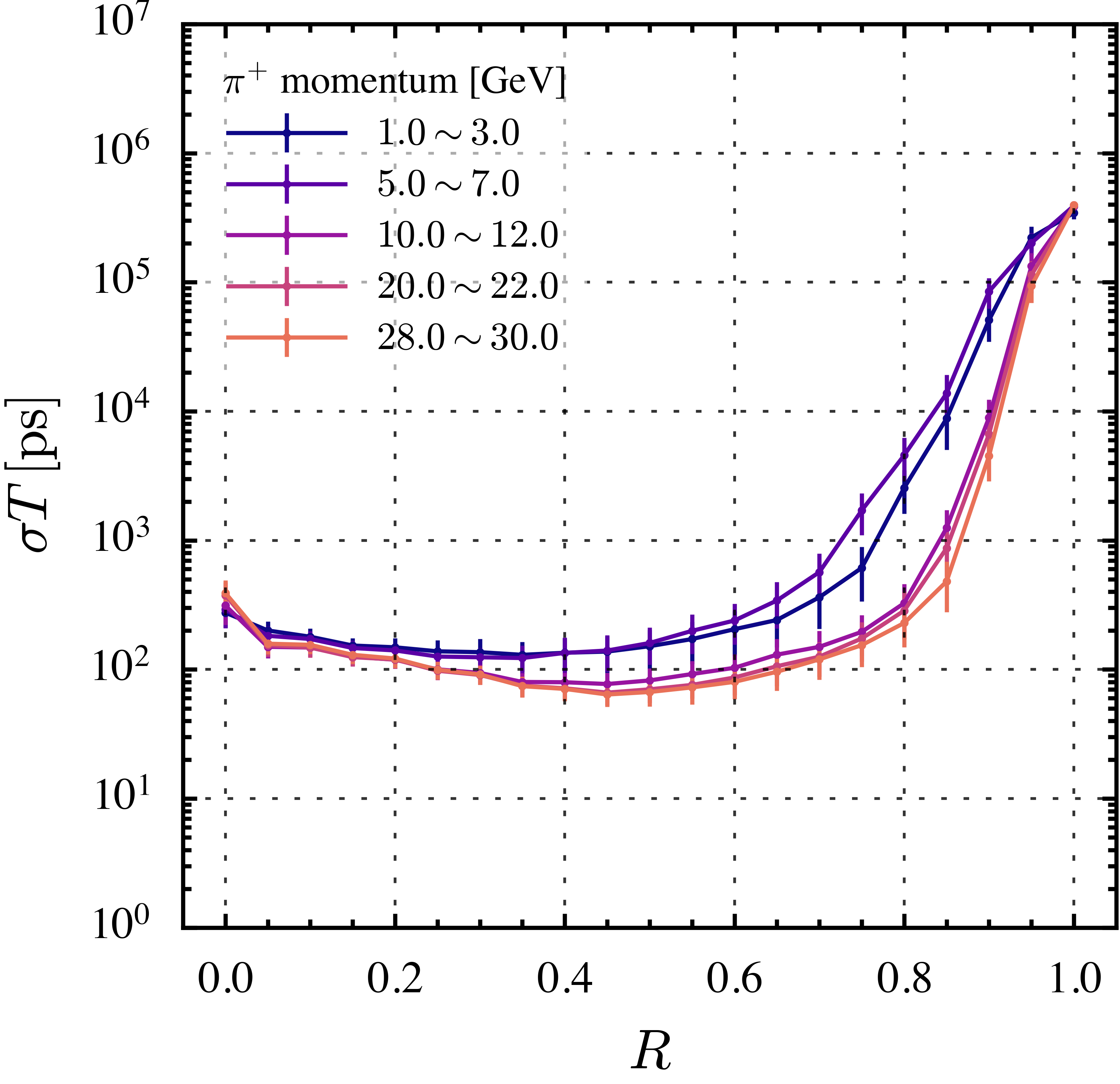}} {\includegraphics[width=0.32\textwidth]{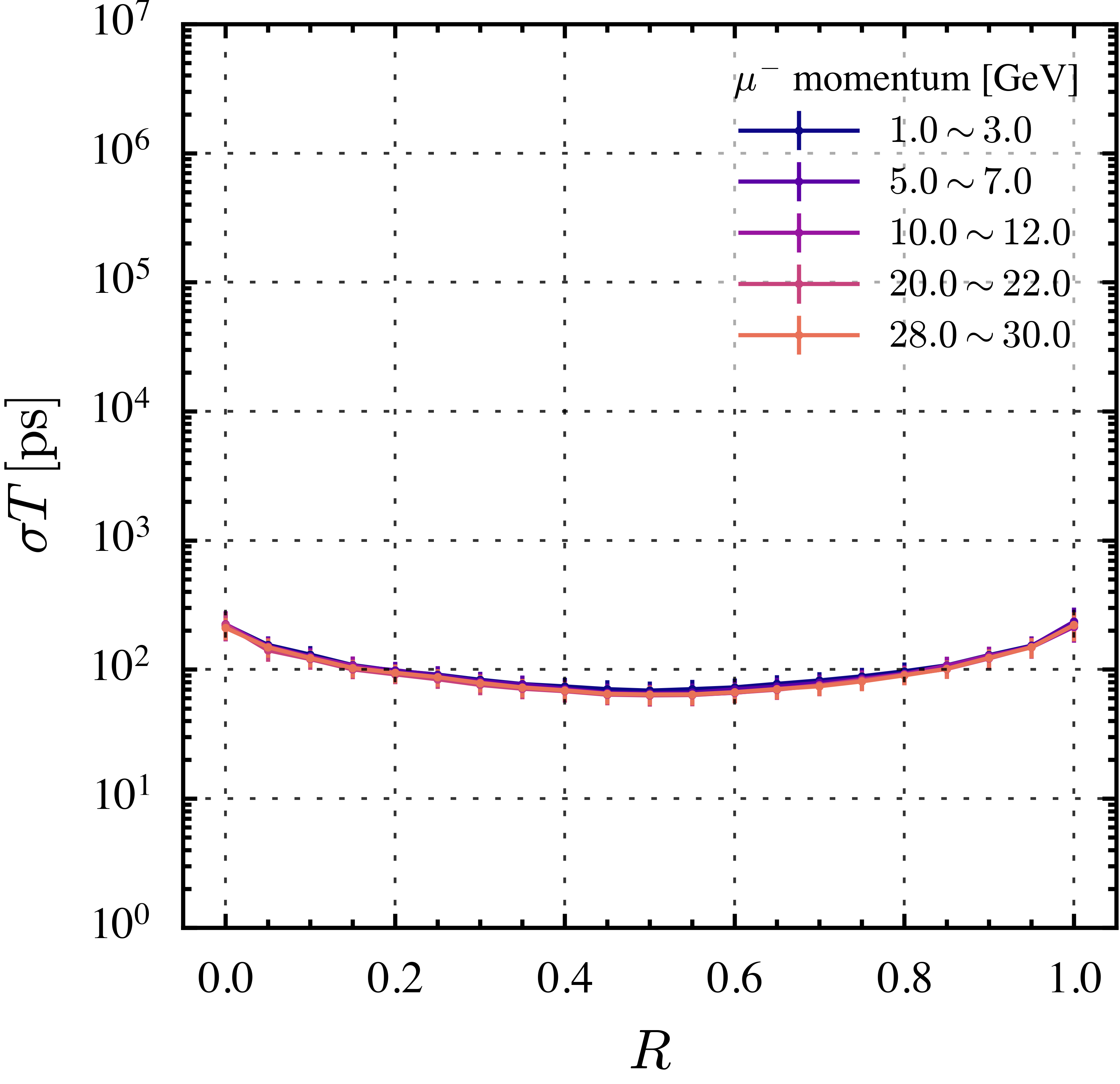}}
  \caption{The shower time resolution in the $\Pphoton$ (left), $\Ppiplus$ (middle), $\Pmuon$ (right) samples as a function of $R$.}
  \label{reso_curve}
\end{figure*}

The input of the algorithm is the projected times of the hits in the shower. A
dedicated clustering algorithm is needed to assign the hits to the shower
corresponding to the originated particle. Therefore, the clustering algorithm
affects the input of the time reconstruction algorithm. In this section, to
first decouple the impact of the clustering algorithm, we quantify the
performance of the algorithm in single-particle events by considering all the
hits in each event as a perfect cluster. The reconstructed time is compared
with the true shower time, where the true time of a shower is defined as the
earliest true projected time of the shower hits.

Fig. \ref{fig:residual} shows the distribution of the time difference between
the reconstructed value and the true value in the 10 GeV $\Ppiplus$ sample.
This distribution highly depends on the parameter $R$. When $R$ is too small or
too large, the resulting residual spectrum shows a large width followed by a
long tail on one side. On the contrary, when $R$ is in an optimal region, the
residual distribution is relatively narrow and symmetrical.

The bias and resolution of time reconstruction are extracted from the residual
distribution to quantify and optimize the algorithm performance. The bias and
resolution are defined as the mean and standard deviation of the residual,
respectively. The error on the resolution is evaluated by $
  \frac{1}{2\sigma}\sqrt{\frac{1}{n}\left(\mu_{4} -
    \frac{n-3}{n-1}\sigma^{4}\right)}$ \cite{rao_linear_2002} where $\mu_4$ and
$\sigma$ are the fourth moment and standard deviation of the residual, and $n$
denotes the number of hits in the bin. The calculated bias and resolution
versus $R$ are shown in Fig. \ref{bias_curve} and \ref{reso_curve}. Except for
muons, the $R$ minimizing the resolution slightly differs from that for
unbiased reconstruction. In more detail, $R$ can be optimized according to the
PID information and energy. The optimal $R$ of hadronic showers is smaller than
the value of the EM shower since the tail of the projected time spectrum of
hadronic showers is much more significant than that of photon and muon.
Moreover, the muon contains few later hits, so this kind of shower corresponds
to an $R \sim 0.5$.

\begin{figure}[htb]
  \centering        		\includegraphics[width=0.43\textwidth]{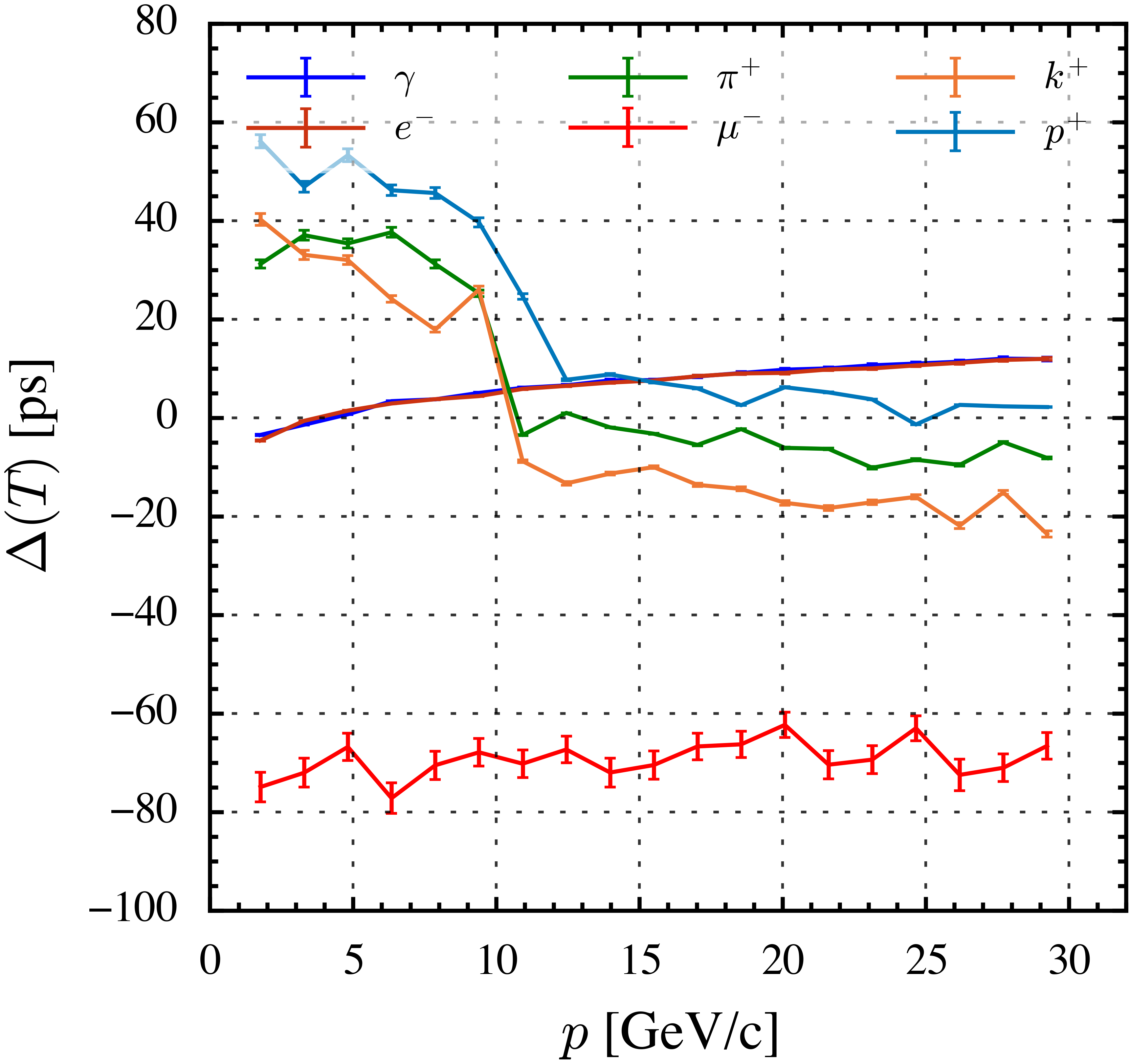}
  \quad
  \includegraphics[width=0.43\textwidth]{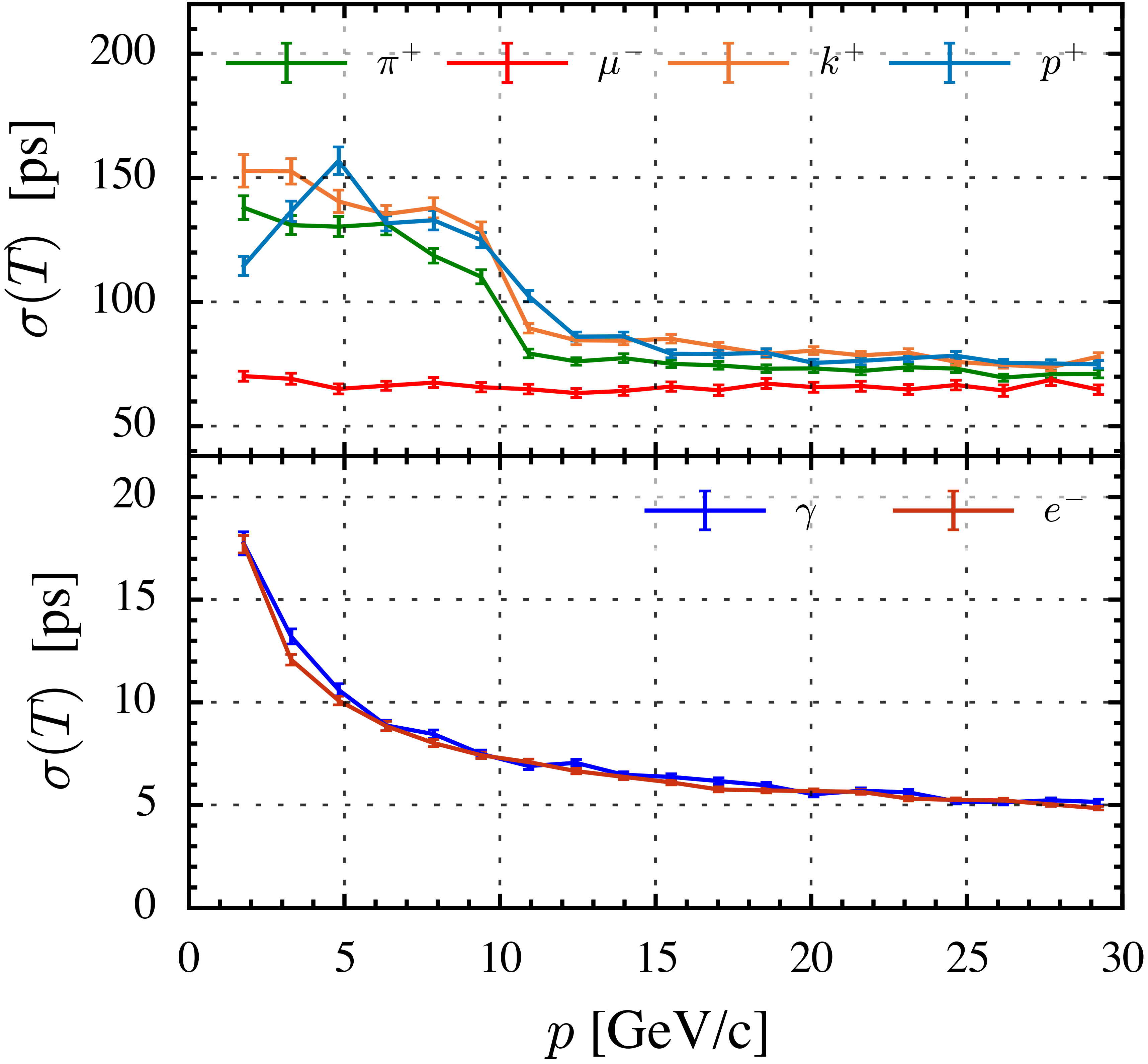}
  \caption{With $R=0.4$, the time reconstruction bias (top) and resolution (bottom) of $\Pelectron$, $\Pmuon$,$\Ppiplus$,$\PKplus$ and $\Pproton$ as a function of the incident momentum.}
  \label{bias_reso_en}
\end{figure}

The dependence of the bias and resolution on the incident momentum of the
particle is shown in Fig. \ref{bias_reso_en}. The value of $R$ is fixed at 0.4.
The time resolution of the EM showers with energy higher than 1 GeV is better
than 20 ps. As the incident momentum increases, the time resolution improves
statistically because of the increasing number of hits in showers. When the
momentum increases above $25 \mathrm{~GeV}$, the time resolution reaches less
than $5 \mathrm{~ps}$. Because the thickness of the ECAL is only
$26~\mathrm{X_0}$, a fraction of the energy of the hadronic particle can not
deposit in ECAL. This fact causes the time resolution of the hadronic showers
to be 80 to 150 ps, worse than that of EM showers. In Fig. \ref{bias_reso_en},
there appears to be a step around 10 GeV. This step exists in the hadronic
particle samples simulated with the physics lists of QGSP\_BERT and
QGSP\_BERT\_HP and is waiting for further exploration. Moreover, as minimum
ionization particles, muons tend to create about one hit of $\sim 1$ MIP per
layer along their trajectory. Consequently, the time resolution of muons is
independent of the incident energy and is nearly 1/5 the intrinsic time
resolution of an individual one MIP hit.

\section{Scaling with changing intrinsic hit time resolution and layers number}\label{sec5}

\begin{figure}[htbp]
  \centering
  \includegraphics[width=0.43\textwidth]{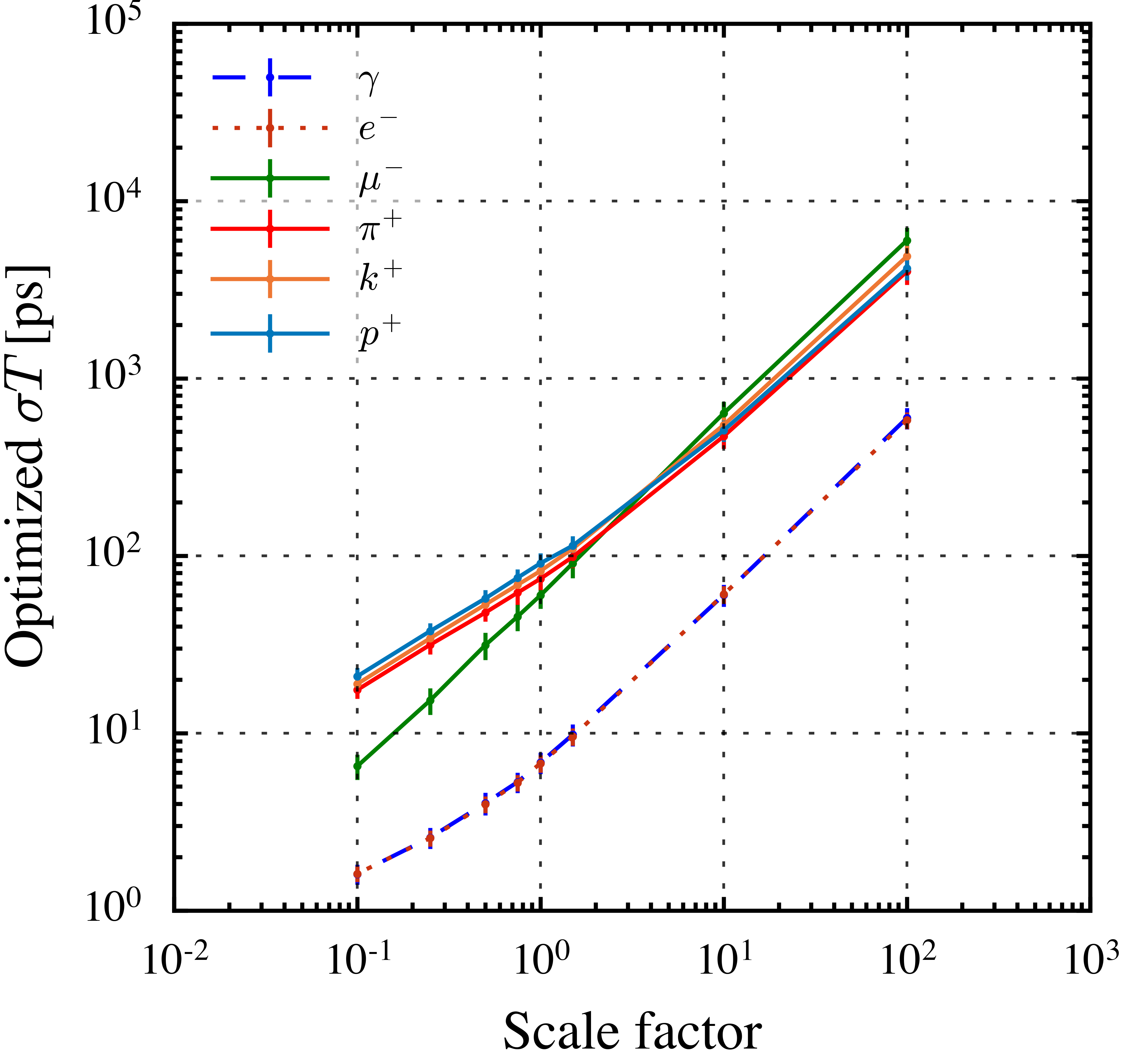}
  \quad
  \includegraphics[width=0.43\textwidth]{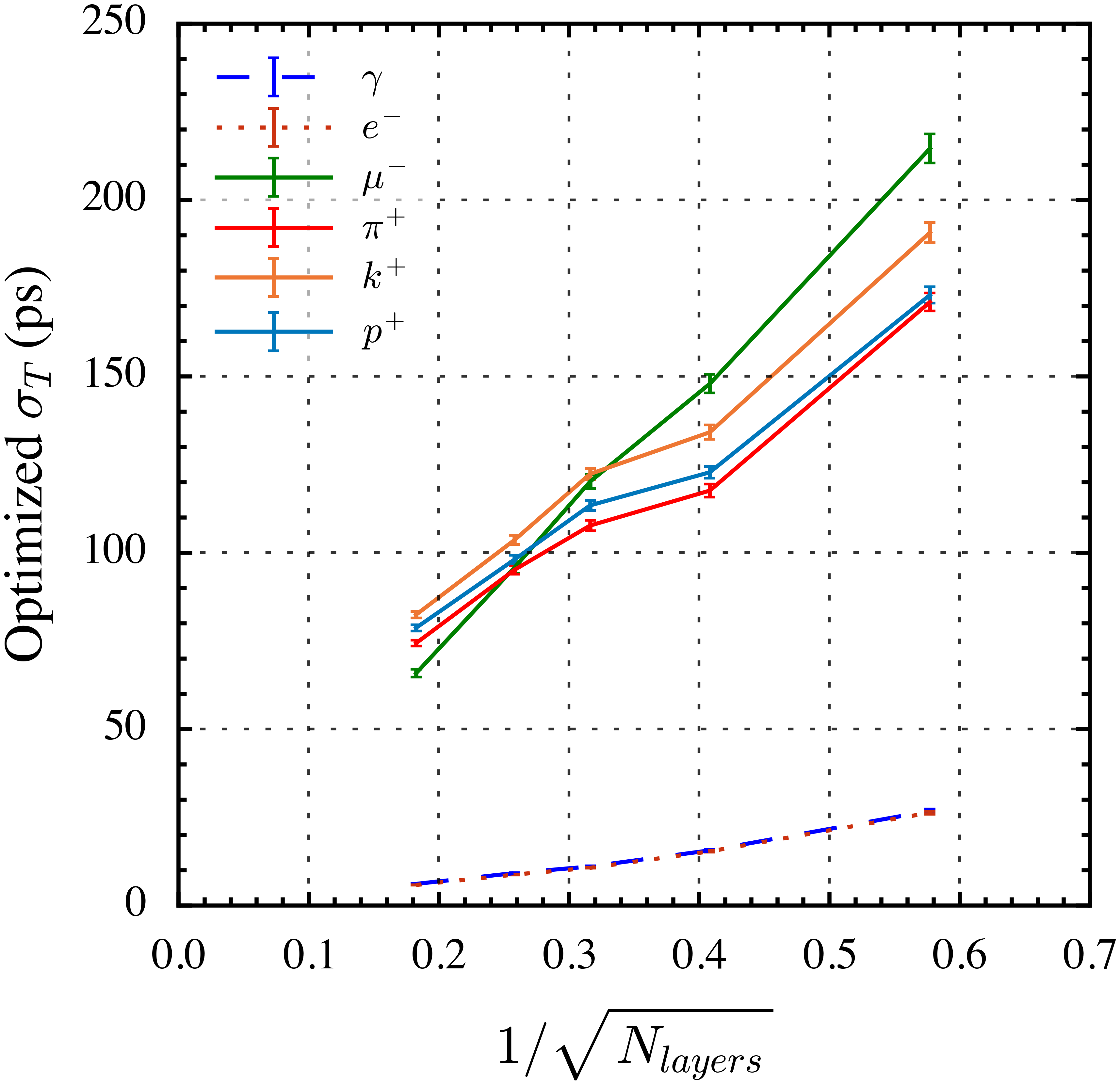}
  \caption{The scaling behavior of the shower time resolution for $10 \sim 15 \mathrm{GeV}$ particles versus the intrinsic hit time resolution (top) and the number of timing layers (bottom). $\alpha$ denotes the scale factor in Eq. \ref{eq:scale_dt}}
  \label{fig:scale}
\end{figure}

\begin{figure}[htbp]
  \centering
  \includegraphics[width=0.43\textwidth]{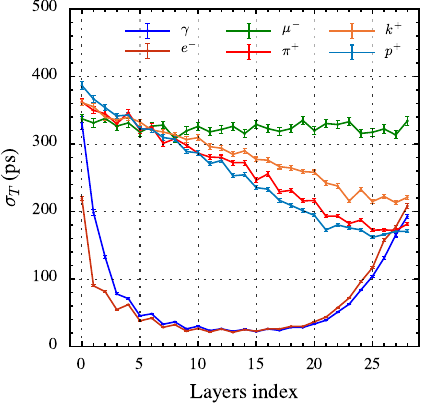}
  \includegraphics[width=0.43\textwidth]{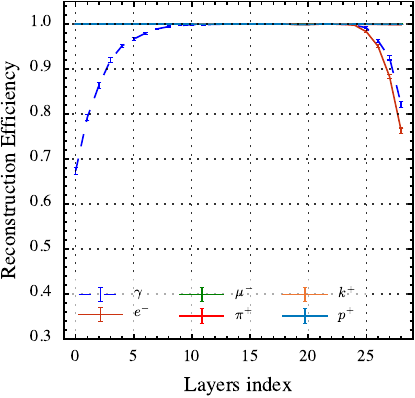}
  \caption{The time resolution (top) and corresponding efficiency (bottom) from a single layer for $25 \sim 30 \mathrm{GeV}$ particles as a function of the layer index.}
  \label{fig:single_layer}
\end{figure}

The cluster ToF performance strongly depends on the intrinsic time resolution
of each individual ECAL channel. In this section, we quantify the dependence of
ToF resolution on the intrinsic hit time resolution by scaling the intrinsic
hit time resolution with a factor $\alpha$,
\begin{equation}
  \label{eq:scale_dt}
  \sigma = \alpha \cdot \sqrt{\left(\frac{A}{E}\right)^2 + C^2}, \left(
  \begin{array}{l}
      A = 380 \;\mathrm{ps \cdot MIP}, \\
      C = 10 \;\mathrm{ps}
    \end{array}
  \right)
\end{equation}
and observing the optimal resolution versus different factors. Furthermore, the
arrangement of the timing readout layers also impacts the ToF reconstruction.
This arrangement should be optimized to balance the detector performance and
building cost. In order to briefly explore the impact of the timing layers
arrangement, we only choose a part of the ECAL layers with equal distance,
perform the time reconstruction using the hits on these chosen layers, and
finally observe the relationship between the ToF resolution and the number of
timing layers. In the case of only several timing layers, the impact of the
layer position is also discussed.

The ToF resolution versus the scaling factor ($\alpha$) of the intrinsic hit
time resolution is shown in Figure \ref{fig:scale}. When the single hit time
resolution is scaled from 100 to 0.1 times the level in Fig.
\ref{g_int_hit_reso}, the cluster time resolution reflects an approximately
linear relationship with the intrinsic hit time resolution.

The shower time resolution as a function of the number of layers is shown in
Fig. \ref{fig:scale}, where the timing layers are arranged at isometric
intervals. The cluster timing performance statistically deteriorates when the
number of timing layers decreases since the input hit times of the algorithm
become fewer and fewer. Fig. \ref{fig:single_layer} shows the impact of the
timing readout arrangement when only one layer provides timing information. EM
shower time resolution reaches the highest when the timing readout is located
on the $10\sim 15$th layer, where the depth corresponds to $ 6\sim 9
  \mathrm{~X_0}$ and the energy deposition of EM showers is more intense. The
optimal time resolution of EM showers is about $\sim 20 \mathrm{~ps}$, which is
consistent with the test beam result of the CMS HGCAL timing layer
\cite{apresyan2016test, apresyan2016investigation}. When the timing readout is
installed on the first or last few layers, the reconstruction performance and
efficiency decrease, as shown in the second plot of Fig.
\ref{fig:single_layer}. For hadronic showers, the time resolution improves with
the timing layer moving to a deeper position. The effect of timing layers
arrangement on the timing of muon is marginal since the energy deposition is
highly uniform. Furthermore, considering the better timing performance arising
from the layers around the shower maximum, the shower timing performance is
hopeful to be improved by installing several dedicated silicon timing layers
with high precision, such as the LGAD \cite{sadrozinski20174d,
  atlas2020technical}, at the key position. From this perspective, further
research and testing about the response of these high-precision timing sensors
on calorimeters will be beneficial.

\begin{table*}[ht]
  \centering
  \caption{The depletion thickness of silicon sensors, assumed noise term coefficient of the intrinsic time resolution (which is the same with the beam test result in Ref ~\cite{akchurin2017timing}), and ToF resolution for photons with transverse momentum of 5 GeV estimated in the setup of CEPC ECAL and the three parts of CE-E along transverse radius.}
  \label{tab:cms_tof_estimate}

  \begin{tabular}{llll}
    \hline
    Radius range (cm)                                                   & $30\sim 70$  & $70\sim 100$ & $100\sim 180$ \\ \hline
    p ($p_{t} = 5 \mathrm{GeV}$)                                        & $23 \sim 54$ & $17 \sim 23$ & $10\sim 17$   \\
    Shower time resolution on CEPC ECAL (ps)                            & $< 5$        & $\sim 6$     & $6 \sim 7$    \\
    Active thickness ($\mathrm{\mu m}$)                                 & 120          & 200          & 300           \\
    Noise term A ($\mathrm{ns} \cdot \mathrm{MIP}$)                     & 0.69         & 0.38         & 0.34          \\
    Variance factor of $\sigma(t)$ contributed by intrinsic $\sigma(t)$ & 1.8          & 1            & 0.9           \\
    Variance factor of $\sigma(t)$ contributed by $N_{layers}$          & 1.07         & 1.07         & 1.07          \\
    Shower time resolution on CMS CE-E (ps)                             & $< 10$       & $\sim 6$     & $~6$          \\ \hline
  \end{tabular}
\end{table*}

With the above observation, we can briefly estimate the timing precision of the
CMS CE-E. The difference in the number of timing layers can only contribute a
variance of 7\% on shower time resolution. On the other hand, the depletion
thickness of silicon sensors is 120 $\mathrm{\mu m}$, 200 $\mathrm{\mu m}$ and
300 $\mathrm{\mu m}$ in the three parts of CE-E corresponding to different
radiation fluence. We assume that the three types of silicon sensors can
provide the same time resolution of the sensors with depletion thickness of 133
$\mathrm{\mu m}$, 211 $\mathrm{\mu m}$ and 285 $\mathrm{\mu m}$ tested in Ref
~\cite{akchurin2017timing}, and the time uncertainty from the distributed
clocks of each channel can be well controlled by calibration algorithms. The
ToF resolution of photons with $p_{T} = 5 \mathrm{~GeV}$ that can be reached in
the three parts of CE-E should be approximately 1.8, 1, and 0.9 times the
resolution estimated in the CEPC ECAL setup, which is listed in Table
\ref{tab:cms_tof_estimate}.

\begin{figure}[ht]
  \centering
  \includegraphics[width=0.43\textwidth]{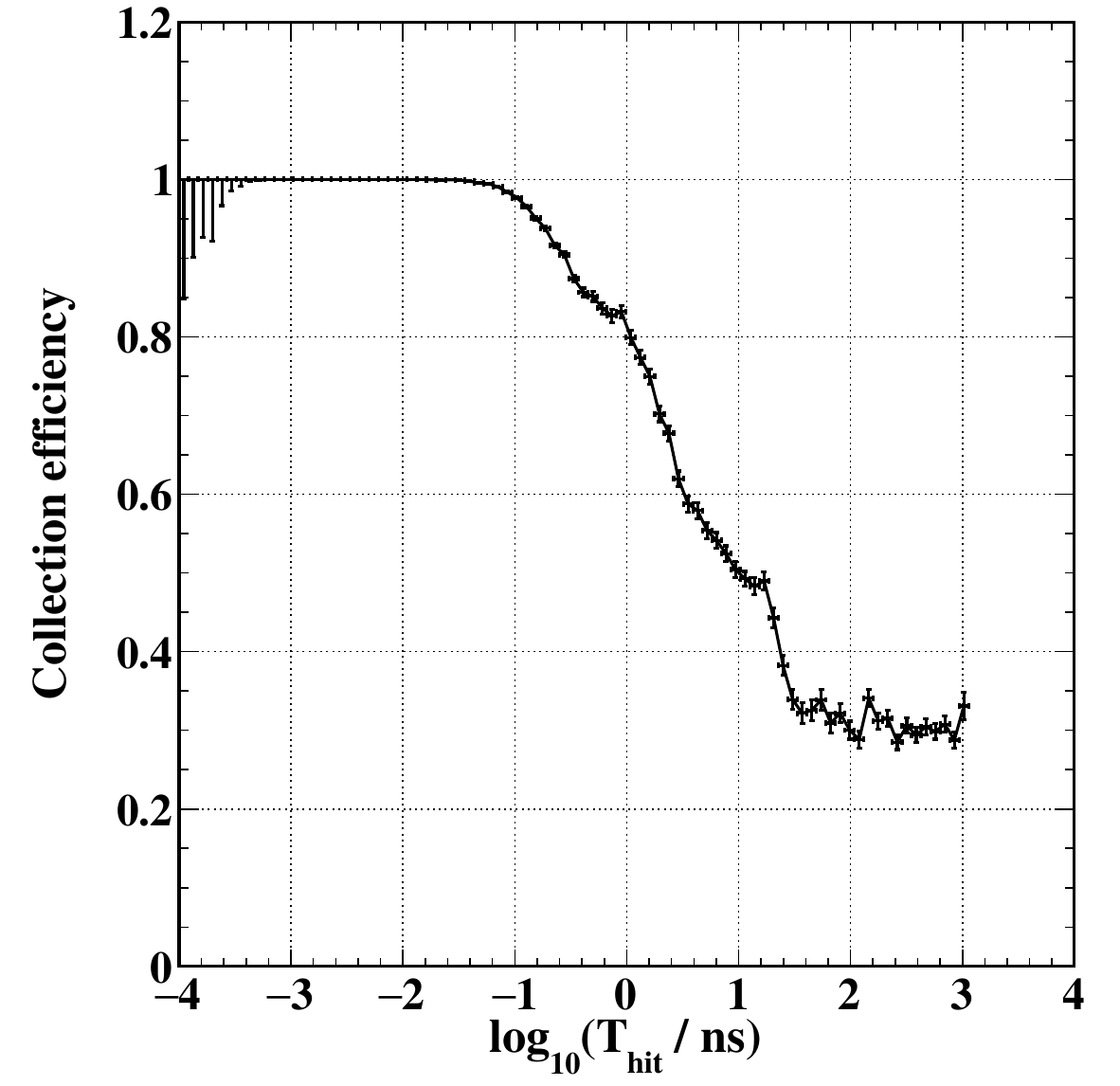}
  \caption{The hit collection efficiency of Arbor versus the true time of ECAL hits, calculated in the 10 GeV $\Ppiplus$ sample.}
  \label{collection_eff}
\end{figure}

\section{Impact of realistic clustering module}\label{sec6}

\begin{figure*}[htbp]
  \begin{minipage}[t]{0.3\textwidth}
    \vspace{0pt}
    \caption{The shower time resolution as a function of $R$ value for perfect clusters and Arbor clusters in the photon (left) sample and the pion (right) samples. The error bars are multiplied by a factor of five for visibility.}
    \label{fig:arbor_r_reso}
  \end{minipage}
  \begin{minipage}[t]{0.7\textwidth}
    \vspace{0pt}
    \centering
    \includegraphics[height=0.41\textwidth]{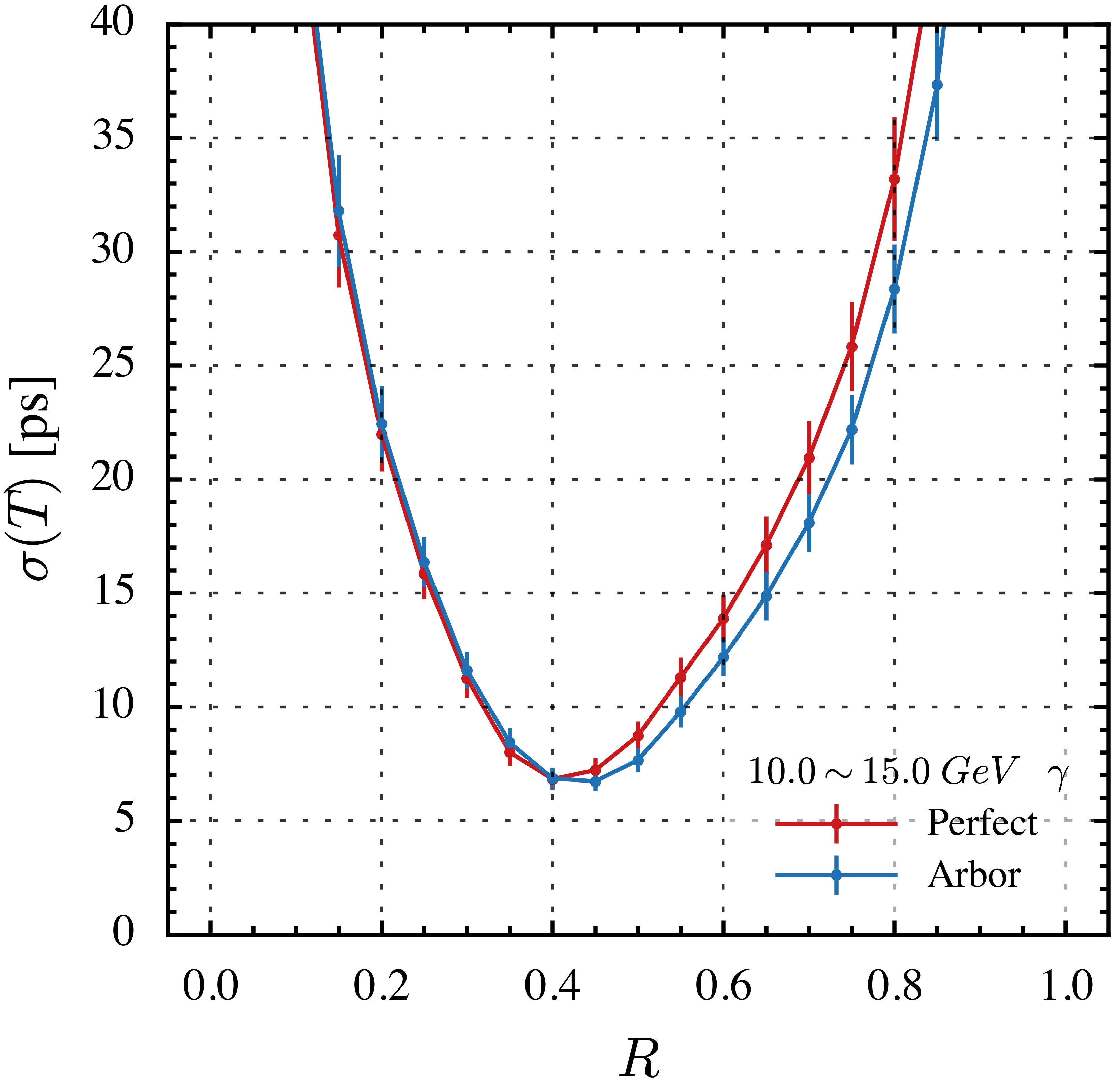}
    \includegraphics[height=0.41\textwidth]{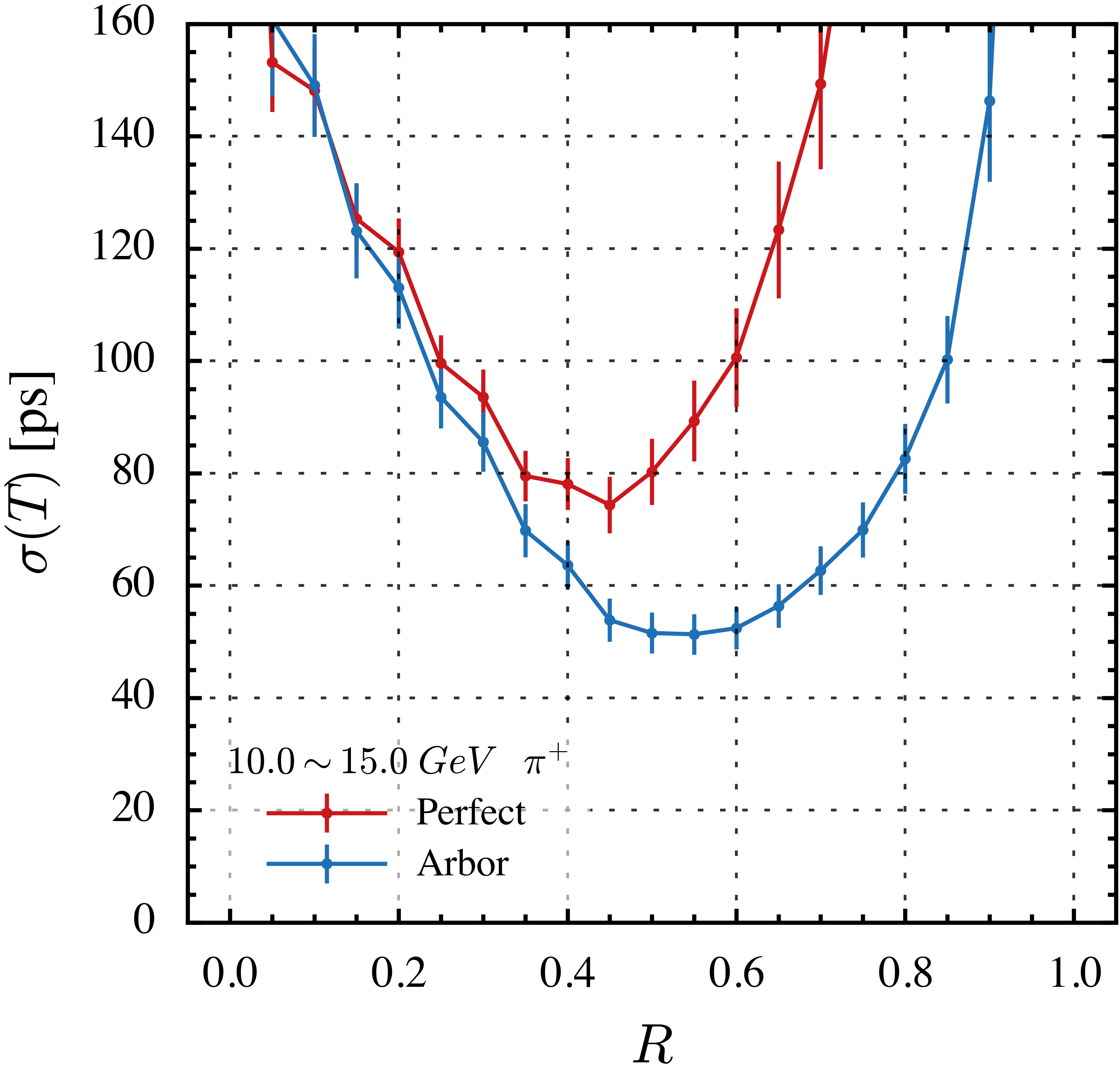}
  \end{minipage}
\end{figure*}

\begin{figure*}[htbp]
  \begin{minipage}[t]{0.3\textwidth}
    \vspace{0pt}
    \caption{The shower time resolution for the Arbor clusters (left) and the time resolution ratio of perfect clusters over Arbor clusters (right) as a function of incident momentum.}
    \label{fig:en_func_osv_pa}
  \end{minipage}
  \begin{minipage}[t]{0.7\textwidth}
    \vspace{0pt}
    \centering
    \includegraphics[width=0.43\textwidth]{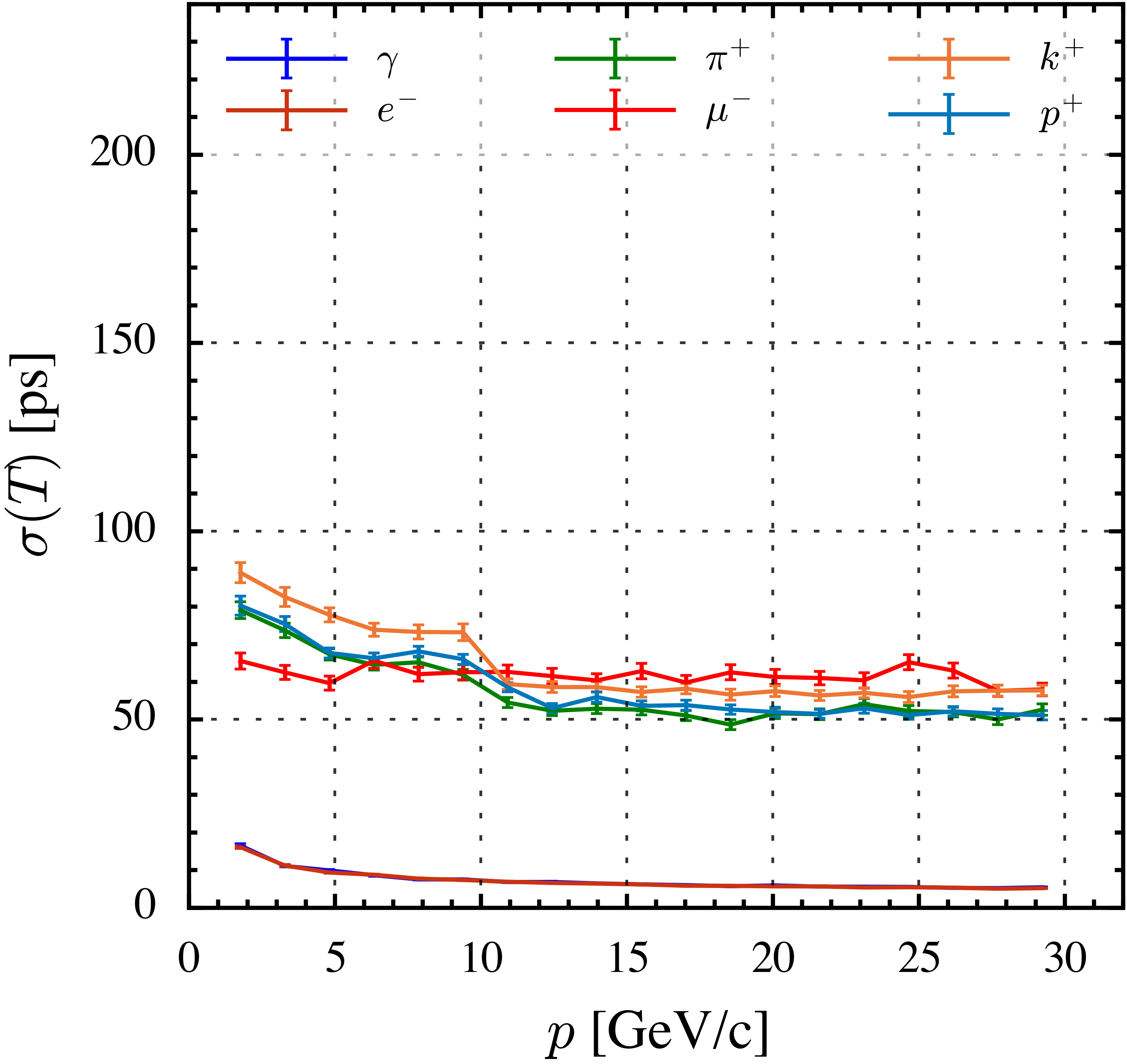}\quad
    \includegraphics[width=0.43\textwidth]{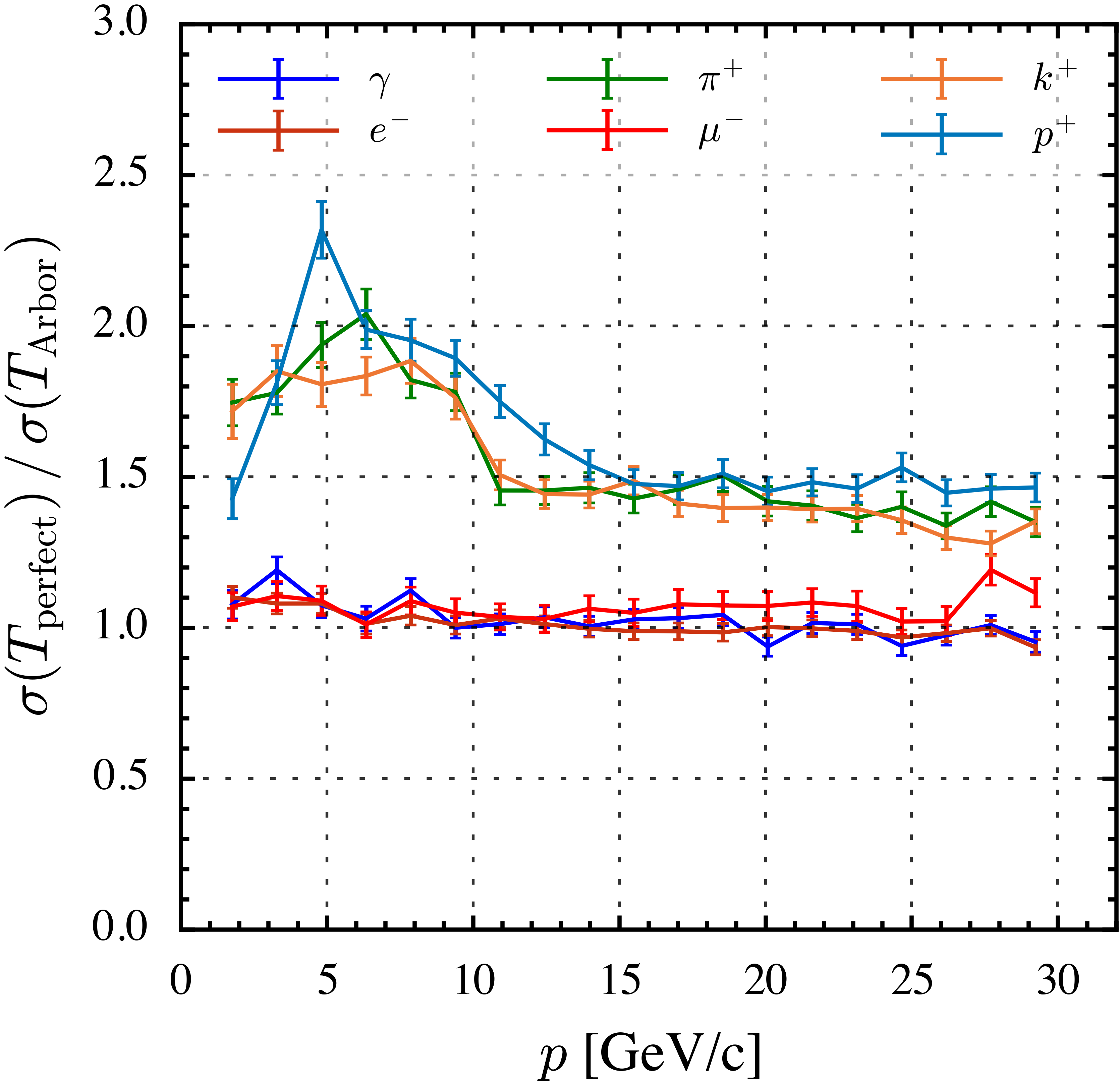}
  \end{minipage}
\end{figure*}

The clustering algorithm is the core part of the PFA, which divides the hits
into clusters corresponding to the shower generated by the final states. In the
ideal case, all the hits can be collected to the clusters that, one by one,
correspond to the incident particles. However, this task is difficult in the
real-world scenario with high particle multiplicity. In fact, various
algorithms use different strategies to decide how to cluster the hits. For
example, the PFA used in the CEPC, Arbor \cite{ruan2014arbor}, selects hits
according to their position and tends to remove the shower hits on the
periphery of the shower. Because of the correlation of the hit time and
position, Arbor has higher collection efficiency for hits in the fast component
than for later hits, as shown in Fig. \ref{collection_eff}. Consequently, the
clustering algorithm impacts the ToF reconstruction.

Fig. \ref{fig:arbor_r_reso} compares the time reconstruction bias and
resolution for the perfect and Arbor clustering algorithms. Because the later
hits are partly removed, the optimal $R$ for the Arbor clusters is slightly
larger than that of the perfect clusters. Fig. \ref{fig:en_func_osv_pa} shows
the ToF resolution of the Arbor clusters and its ratio over the resolution for
the perfect clusters. $R$ is fixed at 0.45 for Arbor clusters and 0.4 for
perfect clusters in this figure. Arbor can improve the time resolution of
$50\sim 90\%$ for hadronic clusters, while the improvement for the EM clusters
is up to $10\%$. In addition, the improvement of the hadronic showers is more
significant when the incident momentum is lower than 10 GeV. The step in Fig.
\ref{bias_reso_en} disappears in the first plot of Fig.
\ref{fig:en_func_osv_pa}, which implies that the step in Fig.
\ref{bias_reso_en} may arise from the hits on the periphery of the shower.

\section{Summary}\label{sec7}


\begin{figure*}[ht]
  \centering
  \subfigure
  {\includegraphics[width=0.32\textwidth]{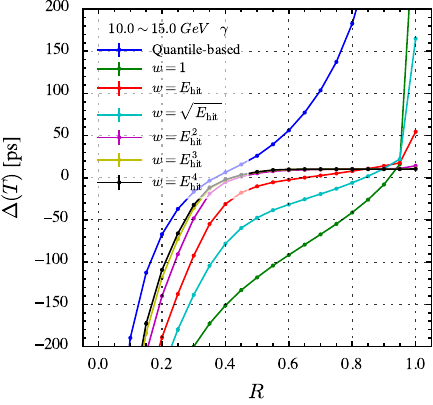}} {\includegraphics[width=0.32\textwidth]{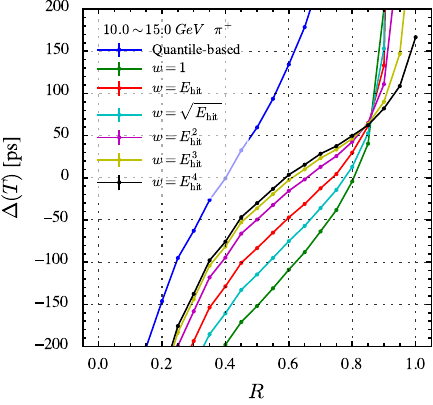}} {\includegraphics[width=0.32\textwidth]{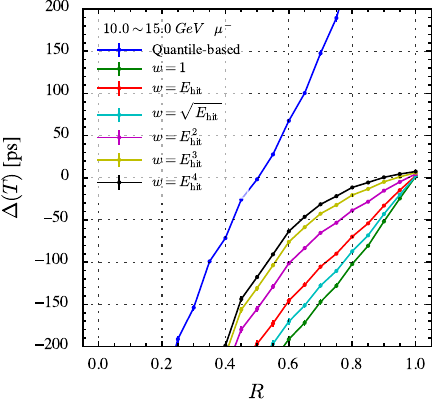}}
  \caption{The shower time reconstruction bias from quantile-based algorithm and average-based algorithms with energy weights of $E_{\mathrm{hit}}^n, ~n = 0,1/2,1,2,3,4$ as a function of $R$.}
  {\includegraphics[width=0.32\textwidth]{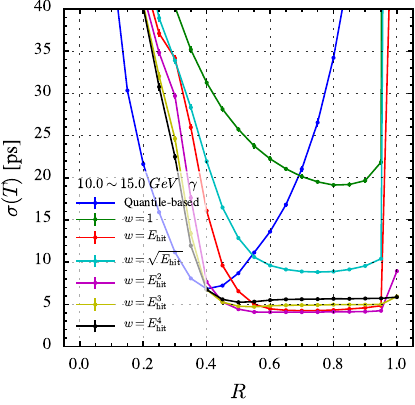}} {\includegraphics[width=0.32\textwidth]{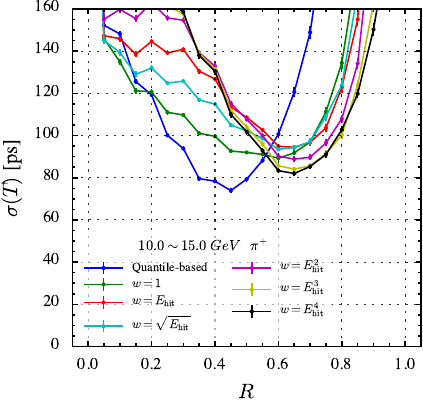}} {\includegraphics[width=0.32\textwidth]{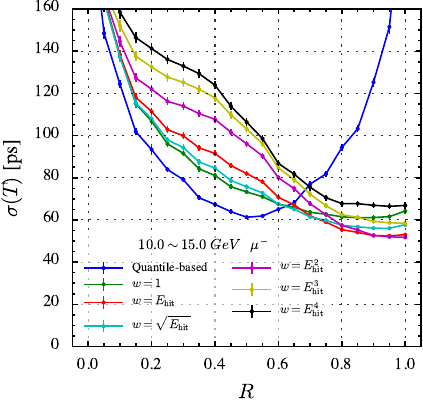}}
  \caption{The shower timing bias(top) and resolution(bottom) from quantile-based algorithm and average-based algorithms with energy weights of $E_{\mathrm{hit}}^n, ~n = 0,1/2,1,2,3,4$ as a function of $R$.}
  \label{osv_kfv_reso}
\end{figure*}


We propose a quantile-based time reconstruction algorithm that extracts the ToF
of particles from shower hits in HGC with typical density of $\mathcal{O}(1
  \sim 10)$ channels per cubic centimeter. The hit time is subtracted by the time
it takes for the light to travel from the IP to the hit position. This
algorithm chooses the quantile of the hit times as the reconstructed shower
time. The time resolution of each channel is parameterized to be $\frac{0.38
    \mathrm{~ns \cdot MIP}}{E}$ $\oplus$ $10 \mathrm{~ps}$ according to the test of
the CMS silicon sensor in Ref.~\cite{akchurin2017timing}. We expect that a time
resolution of 5 to 20 ps (80 to 160 ps) can be achieved for EM (hadronic)
showers on the CEPC ECAL.

The presented algorithm and several alternative strategies based on the average
of the hit times with energy weighting are compared. The average with $E^2$
weighting can improve the time resolution of EM (hadronic) showers from $20
  \mathrm{~ps}$ ($90 \mathrm{~ps}$) to $4 \mathrm{~ps}$ ($85 \mathrm{~ps}$),
comparable with the quantile-based algorithm.

We investigate three relevant factors for the ToF measurement, the intrinsic
hit time resolution, the number of layers, and the time clustering algorithm.
First, we observe an approximately linear dependence between the cluster ToF
resolution and the intrinsic time resolution of each channel. Second, the time
resolution of the showers statistically improves with the number of calorimeter
timing layers if these timing readouts are uniformly installed in the ECAL.
When the number of timing layers is small, optimizing the layout of timing
layers is favorable. The time resolution of EM showers can be significantly
improved by installing the layer in $6 \sim 9 \mathrm{X_0}$, while the last ten
layers are more advantageous for timing hadronic showers. Thirdly, a specific
clustering algorithm affects the timing performance because the hit time is
correlated with the hit position. For example, Arbor leads to an improvement of
up to $<10\%$ ($40 \sim 90\%$) for EM (hadronic) showers, compared to an ideal
clustering module.

With the understanding above, we evaluate the timing performance of the
electromagnetic compartment of the CMS endcap calorimeter. If the intrinsic
time resolution of each channel is similar to the result of
Ref.~\cite{akchurin2017timing}, we expect that this calorimeter can provide 6
to 10 ns time resolution for photons with a transverse momentum of 5 GeV. This
precision is beneficial for pile-up mitigation.

In this work, the distributed clock is assumed to be well synchronized by
hardware technology and calibration algorithms. The current silicon sensors
have high precision timing capability for hits with energy of hundreds MIPs but
relatively low time resolution for those with only several MIPs. In addition,
there are still many interesting patterns in the true time-energy spectrum of
the showers. As HGC timing performance improves to picosecond level, these
patterns will increasingly affect relative measurements. Therefore, it is
valuable to model the timing information of the showers with high precision in
Monte Carlo simulations and analyse the patterns at the picosecond level.

\begin{acknowledgements}
  The authors thank Jianbei Liu, Yong Liu, Huaqiao ZHANG and Jean-Claude Brient for their help in interpreting the timing readout technique. We also thank Chengdong FU and Dan YU for their assistance with the simulation software, and Xuewei JIA for fruitful discussions and language support in this work.
  This project is supported by the International Partnership Program of Chinese Academy of Sciences (Grant No. 113111KYSB20190030), the Innovative Scientific Program of Institute of High Energy Physics.
\end{acknowledgements}

\appendix

\section{Average based strategy}\label{secA1}

This section compares the proposed time reconstruction algorithm with several
alternative average-based strategies that average over the hit times with
different energy weights after removing a part of the shower hit relatively
later in the time spectrum.

The average-based algorithms first sort the recorded hits in ascending order
according to the hit time. Secondly, the energy-weighted average time of the
first $(R^{\prime}\cdot N_{hits})$ hits is calculated as the result.
$R^{\prime}$ is also an ad hoc ratio similar to $R$ in the quantile-based
algorithm. The quantile-based algorithm essentially removes a part of the
latest hits in the shower, where the fraction of removed hits is $2R$. Then the
$R\cdot N_{\mathrm{hits}}$-th hit time is equivalent to the median of the
remaining hit times. In this study, five hit energy weighting models,
$E_{\mathrm{hit}}^{n} ~(n = 0, 1/2, 1, 2, 3)$, are considered.

The quantification in Sec. \ref{sec4} is applied to the average-based algorithm
in photon, charged pion, and muon samples. As shown in Fig. \ref{osv_kfv_reso},
the performance of the average-based strategy can be significantly improved by
energy weighting. The time resolution of EM showers from $E^2$ weighted average
strategy reaches $\sim 4 \mathrm{~ps}$, $\sim 5$ times better than the result
of the unweighted average strategy and higher than the quantile-based algorithm
by $\sim 2 \mathrm{~ps}$. For the hadronic showers and MIP, the impact of
energy weighting on the time resolution is relatively small.

\bibliographystyle{spphys} \bibliography{bibliography}

\end{document}